\documentclass[aps,prd,preprint,amsmath,amssymb]{revtex4}
\usepackage{graphicx}% Include figure files
\usepackage{dcolumn}% Align table columns on decimal point
\usepackage{bm}% bold math
\newcommand{\be}{\begin{equation*}}
\newcommand{\ee}{\end{equation*}}
\newcommand{\bea}{\begin{eqnarray}}
\newcommand{\eea}{\end{eqnarray}}
\newcommand{\bean}{\begin{eqnarray*}}
\newcommand{\eean}{\end{eqnarray*}}
\newcommand{\kp}{{$K^{-} p \to \Lambda \pi^- \pi^+ $}}

\begin{document}

\title{Further evidence for the $\Sigma^{*}$ resonance with $J^P=1/2^-$ around 1380 MeV}

\author{
Jia-Jun Wu$^{1}$, S.~Dulat$^{2,3}$ and B.~S.~Zou$^{1,3}$ \\
$^1$ Institute of High Energy Physics, CAS, P.O.Box 918(4), Beijing 100049, China\\
$^2$ School of Physics Science and Technology, Xinjiang
University, Urumqi, 830046, China\\
$^3$ Theoretical Physics Center for Science Facilities, CAS,
Beijing 100049, China}

\date{September 7, 2009}

\begin{abstract}
The unquenched quark models predict the new particle $\Sigma^*$ with
spin parity $J^P=1/2^-$ and its mass is around the well established
$\Sigma^*(1385)$ with $J^P=3/2^+$. Here by using the effective
Lagrangian approach we study \kp\ reaction at the range of
$\Lambda^{*}(1520)$ peak, comparing the resulting total cross
section, and $\pi^+\pi^-$, $\Lambda\pi^+$, $\Lambda\pi^-$ invariant
squared mass distributions for various incident $K^-$ momenta, as
well as the production angular distribution of the $\Lambda$ with
the data from the Lawrence Berkeley Laboratory 25-inch hydrogen
bubble chamber, we find that, apart from the existing resonance
$\Sigma^{*}(1385)$ with $J^P=3/2^+$, there is a strong evidence for
the existence of the new resonance $\Sigma^{*}$ with $J^P=1/2^-$
around 1380 MeV. Higher statistic data on relevant reactions are
needed to clarify the situation.
\end{abstract}

\pacs{14.20.Gk, 13.30.Eg, 13.75.Jz}

\maketitle

\section{Introduction}
\label{s1}

The classical constituent quark models are based on the assumption
of three constituent quarks inside each baryon. They are very
successful for the spatial ground state of baryons, but have serious
problems for the predictions of baryon excitation states. The lowest
excitation of baryons is expected to be the orbital angular momentum
$L=1$ excitation of a quark, resulting to spin-parity $1/2^-$. The
$N^*(1535)$, $\Lambda^*(1405)$ and $\Sigma^*(1620)$ are the lowest
$1/2^-$ baryons from many experiments~\cite{pdg}. There is a
question that why the mass of $\Lambda^*(1405)$  is much less than
$N^*(1535)$. It is very difficult to explain this problem in the
classical $3$-quark models, because the $\Lambda^*(1405)$ with
$(uds)$-quarks is obviously expected to be heavier than $N^*(1535)$
with $(uud)$-quarks. Another problem is about the $\bar{d}/\bar{u}$
asymmetry in the proton with the number of $\bar d$ more than $\bar
u$ by an amount $\bar d-\bar u\approx 0.12$~\cite{Garvey}. If one
wants to solve these problems, one should put the $q\bar{q}$
components in the baryons. The unquenched models give the good
explanation to these problems. For example, in the penta-quark
models~\cite{Helminen,zhu,Zou-Nstar07}, the mass of $N^*(1535)$ with
mainly a $[ud][us]\bar s$ state is heavier than $\Lambda^*(1405)$
with mainly a $[ud][sq]\bar q$ state with $q\bar q=(u\bar u+d\bar
d)/\sqrt{2}$. The 5-quark models may play an important role in the
baryon spectroscopy.

These unquenched models give many new predictions besides the
properties of $\Lambda^*(1405)$ and $N^*(1535)$. In fact, the
penta-quark models~\cite{Helminen,zhu} show a new physical picture
for the baryonic excitation. The lowest excitation is $J^P=1/2^-$ in
the $qqqq\bar{q}$ model, and there are two new particles
$\Sigma^{*}(1360-1405)$ and $\Xi^{*}(1520)$ which are absent in the
$qqq$ model. The meson cloud model~\cite{jido} predicts them to be
non-resonant broad structures. These new predictions are all very
different from the results of the classical quenched quark models,
so it needs to be checked by experiments.

Possible existence of such new $\Sigma^*(1/2^-)$ structure in
$J/\psi$ decays was pointed out earlier~\cite{Zou-Charm06} and is
going to be investigated by the starting BES3
experiment~\cite{BES3-yb}, and we also re-examined the old data of
$K^-p\to\Lambda\pi^+\pi^-$ reaction at $Plab_{(K^{-})}=1.0-1.8$ GeV
to find some evidence for its existence~\cite{kp}. In this paper by
using the results from the fit of experimental data in the
Ref.~\cite{kp}, we show further evidence for the existence of such
$\Sigma^*(1/2^-)$ in the $K^-p\to \Lambda^{*}(1520)\to \Sigma^*\pi
\to\Lambda\pi^+\pi^-$ reaction at $Plab_{(K^{-})}=0.25 - 0.60$ GeV,
with a very clear peak of $\Lambda^{*}(1520)$ in the energy
dependence of the total cross section~\cite{oset,exp}.

In the next section, we present the formalism and ingredients
for the study of the \kp\  reaction by including various
Feynman diagrams. In the last section, our numerical
results, comparision with the experimental data, and conclusions
are given.

\section{Formalism and ingredients}
\label{s2}

\begin{figure}[htbp] \vspace{-0.cm}
\begin{center}
\includegraphics[width=0.9\columnwidth]{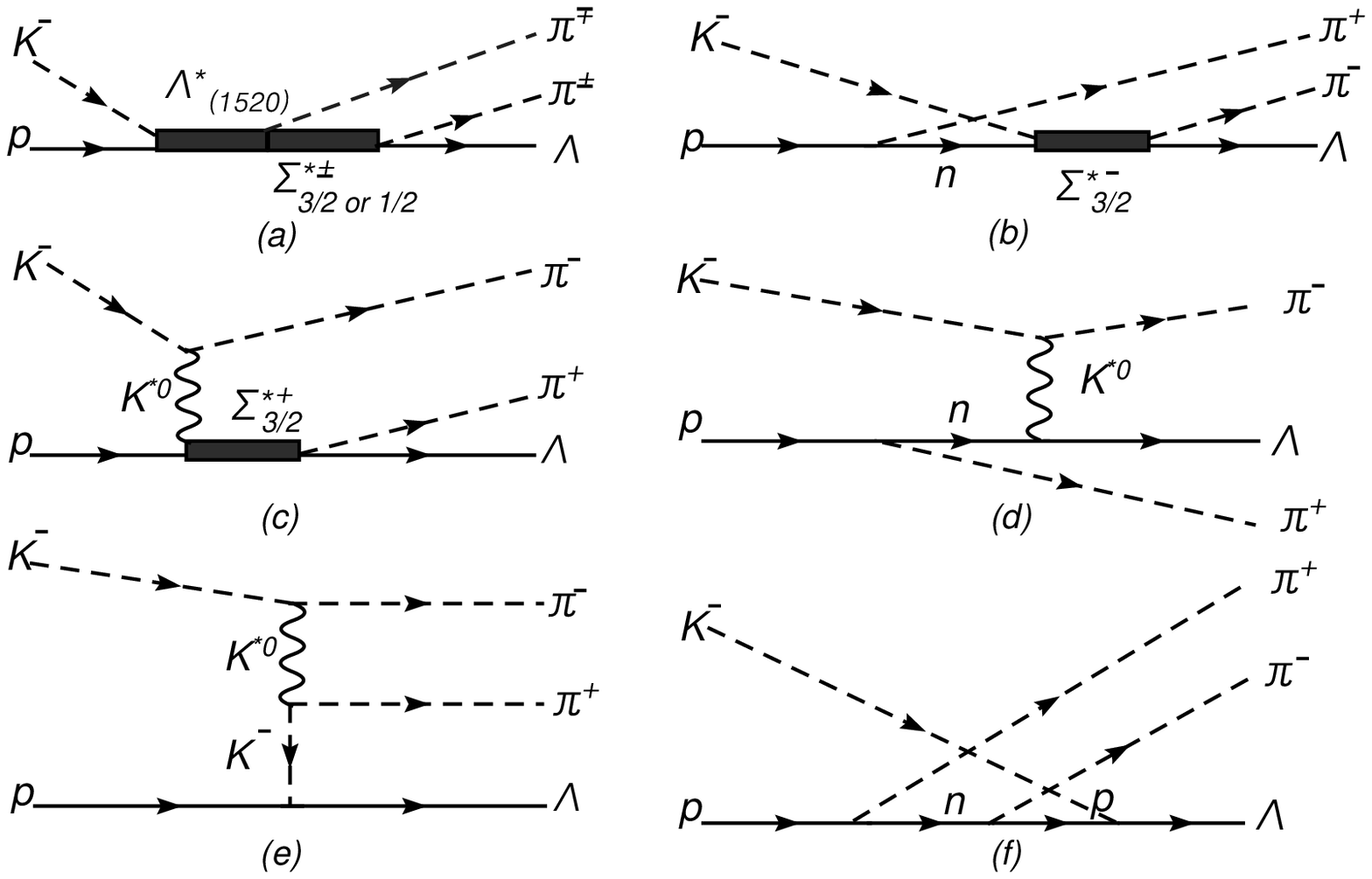}
\caption{The Feynman diagrams of the \kp\  reaction.} \label{fe}
\end{center}
\end{figure}

In this section we present the formalism and ingredients for the
analysis of
\begin{equation}\label{kp}
K^{-} p \to \Lambda \pi^- \pi^+
\end{equation}
in the energy region around the $\Lambda^*(1520)$. First, the
corresponding Feynman diagrams, s-channel $\Lambda^\star(1520)$
exchange diagram (a), u-channel $n$ exchange diagram (b),  t-channel
$K^{*0}$ exchange diagram (c), t-channel $K^{*0}$ and $n$ exchange
diagram (d), t-channel $K^{*0}$ and $K^-$ exchange diagram (e),  and
u-channel $n$ and $p$ exchange diagram (f), for the reaction
(\ref{kp}) are depicted in Fig. \ref{fe}.

Besides we give the effective Lagrangian densities for describing
the interaction vertices in Fig. \ref{fe}. They can be written as

\begin{eqnarray}
{\cal L}_{\Lambda^{*} K N}&=&g_{\Lambda^{*}K N}
\bar{\Lambda}^{*\nu}\gamma_{5}\gamma^{\mu}N\partial_{\nu}\partial_{\mu}\bar{K}+h.c.,\\
{\cal L}_{\Lambda^{*} \Sigma^{*}_{3/2} \pi}&=&g_{\Lambda^{*}
\Sigma^{*}_{3/2} \pi}
\bar{\Lambda}^{*}_{\mu}{\Sigma^{*\mu}_{3/2}}\pi+h.c.,\\
{\cal L}_{\Sigma^{*}_{3/2} \Lambda \pi}&=&g_{\Sigma^{*}_{3/2}
\Lambda \pi}
\bar{\Lambda}{\Sigma^{*\mu}_{3/2}}\partial_{\mu}\pi+h.c.,\\
{\cal L}_{\Lambda^{*} \Sigma^{*}_{1/2} \pi}&=&g_{\Lambda^{*}
\Sigma^{*}_{1/2} \pi}
\bar{\Lambda}^{*}_{\mu}\Sigma^{*}_{1/2}\partial^{\mu}\vec{\pi}\cdot\vec{\tau} +h.c.,\\
{\cal L}_{\Sigma^{*}_{1/2}\Lambda
\pi}&=&g_{\Sigma^{*}_{1/2}\Lambda \pi}
\bar{\Lambda} \Sigma^{*}_{1/2}\vec{\pi}\cdot\vec{\tau}+h.c.,\\
{\cal
L}_{\Sigma^{*}_{3/2}KN}&=&\frac{g_{\Sigma^{*}_{3/2}KN}}{m_{K}}
\vec{\bar{\Sigma}}^{*}_{{3/2}\mu}\cdot\vec{\tau}N\partial^{\mu}\bar{K}+h.c.,\\
{\cal L}_{NN\pi
}&=&\frac{g_{NN\pi}}{2m_{N}}\bar{N}\gamma_{5}\gamma_{\mu}
N\partial^{\mu}\vec{\pi}\cdot\vec{\tau}+h.c.,\\
{\cal L}_{K^{*}K\pi
}&=&g_{K^{*}K\pi}K^{*\mu}(\vec{\pi}\cdot\vec{\tau}\partial_{\mu}K-K\partial_{\mu}\vec{\pi}\cdot\vec{\tau})+h.c.,\\
{\cal L}_{\Sigma^{*}_{3/2}K^{*}N
}&=&\frac{g_{\Sigma^{*}_{3/2}K^{*}N}}{2m_{N}}\vec{\bar{\Sigma}}^{*}_{{3/2}\mu}\cdot\vec{\tau}
\gamma_{\nu}
\gamma_{5}N(\partial^{\nu}\bar{K}^{*\mu}-\partial^{\mu}\bar{K}^{*\nu})+h.c.,\\
{\cal L}_{\Lambda K^{*} N}&=& \bar{\Lambda}(g_{\Lambda K^{*}
N}\gamma_{\mu}+\frac{f_{\Lambda K^{*} N}}{2m_{\Lambda}}
\sigma_{\mu\nu}\partial^{\nu})\bar{K}^{*\mu}N+h.c.,\\
{\cal L}_{\Lambda K N}&=&\frac{g_{\Lambda K N}}{2m_{\Lambda}}
\bar{\Lambda}\gamma_{5}\gamma_{\mu}N\partial^{\mu}\bar{K}+h.c..
\label{mnn}
\end{eqnarray}
Here $m_K$, $m_N$ and $m_\Lambda$ are the kaon, nucleon and
$\Lambda$ masses; $\Sigma^*_{3/2\mu}$ and $\Lambda^*_\mu$ are
Rarita-Schwinger fields for $\Sigma^*(1385)$ and $\Lambda^*(1520)$
of spin-3/2 particles; $\Sigma^*_{1/2}$, $N$ and $\Lambda$ are the
spin-half fields for the $\Sigma^*(1380)$,  N(938) and
$\Lambda(1115)$ particles; $\pi$ and $K$ are scalar fields for the
pion and kaon; $\vec{\tau}$ is a usual isospin-$1/2$ Pauli matrix
operator; the relevant interaction coupling constants, obtained by
using the above effective Lagrangians to fit relevant decay widths
or from literature, are all listed in Table~\ref{tab1}.

\begin{table}[ht]
\begin{tabular}{ c c c c c c c }
\hline \hline R\ \     & $\Gamma_{R}(GeV)$\ & Decay mode\  &
Branching ratios\
& $g^{2}/4\pi(f^{2}/4\pi)$ \\
\hline
 $\Lambda^{*}(1520)$    &  0.0156  & $NK$                                 & 0.45           & 11.88      \\
                        &          & $\Sigma\pi$                          & 0.42           & 7.38         \\
                        &          & $\Sigma^*_{3/2}\pi\to\Lambda\pi\pi$  & 0.11~\cite{exp} & 0.56 (a)     \\
                        &          & $\Sigma^*_{1/2}\pi\to\Lambda\pi\pi$  & 0.11~\cite{exp} & 3.57 (b)     \\
 $K^{*}$                & 0.0508   & $K\pi$                               & 0.9976         & 2.52         \\
 $\Sigma^{*}_{3/2}$     & 0.0358   & $\Lambda\pi$                         & 0.87           & 6.68         \\
                        &          & $K^{*}N$                             &                & 2.39~\cite{Nak}  \\
                        &          & $KN$                                 &                & 0.83~\cite{Nak}  \\
 $\Lambda$              &          & $K^{*}N$                             &                & 1.588 (5.175)~\cite{julich}\\
                        &          & $KN$                                 &                & 3.506~\cite{julich} \\
 $N$                    &          & $N\pi$                               &                & 14.4~\cite{ouyang} \\
 \hline \hline
\end{tabular}

\caption{Parameters used in our calculation. Widths and branching
ratios are from PDG~\cite{pdg}; the mass and width of
$\Sigma^{*}_{1/2}$ are $1.3813$ GeV and $0.1186$ GeV,
respectively~\cite{kp}; for (a) and (b) we use
$(g_{\Lambda^{*}\Sigma^{*}\pi}g_{\Sigma^{*}\Lambda\pi})^{2}/(4\pi)^{2}$,
while  assuming that all $\Lambda\pi\pi$ come from $\Sigma^{*}\pi$
in the $\Lambda^{*}\to\Lambda\pi\pi$ reaction.}\label{tab1}
\end{table}

Furthermore, we need also the propagators of resonant particles to calculate Feynman diagrams. For
the $K$ and $K^{*}$ mesons, the propagators are:
\begin{eqnarray}
G_{K(q)}&=&\frac{1}{q^{2}-m_{K}^{2}},\\
G_{K^{*}(q)}&=&\frac{-g_{\mu\nu}+q^{\mu}q^{\nu}/m^{2}_{K^{*}}}
{q^{2}-m_{K^{*}}^{2}+i\Gamma_{K^{*}}m_{K^{*}}}. \label{bp1}
\end{eqnarray}
For the spin-1/2 and spin-3/2 baryon resonances the propagators can
be written as \cite{ouyang}:
\begin{eqnarray}
G^{\frac{1}{2}}_{R(q)}&=&\frac{(\not\! p + m)}{q^{2}-m_{R}^{2}+im_{R}\Gamma_{R}},\\
G^{\frac{3}{2}}_{R(q)}&=&\frac{(\not\! p + m)}{q^{2}-m_{R}^{2}
+im_{R}\Gamma_{R}}\left(-g_{\mu\nu}+\frac{1}{3}\gamma_{\mu}\gamma_{\nu}
+
\frac{2}{3}\frac{q_{\mu}q_{\nu}}{m_{R}^{2}}+\frac{1}{3m_{R}}(\gamma_{\mu}q_{\nu}-\gamma_{\nu}q_{\mu})\right).
\label{bp3}
\end{eqnarray}
The role of $\Lambda^{*}(1520)$ is very important, thus we take into
account that the width $\Gamma_{\Lambda^{*}(1520)}$ of the
$\Lambda^{*}(1520)$ is dependent on its four-momentum squared, and
by straightforward calculation we obtain the following expression
for the $\Gamma_{\Lambda^{*}(1520)}$
\begin{equation}
\Gamma_{\Lambda^{*}(1520)(s)}= \Gamma_{\Lambda^{*}NK(s)}+\Gamma_{\Lambda^{*}\Sigma\pi(s)}
+\Gamma_{\Lambda^{*}\Lambda\pi\pi(s)}+\Gamma_{0},\\
\end{equation}
where
\begin{eqnarray}
\Gamma_{\Lambda^{*}NK(s)}&=&\frac{g^{2}_{\Lambda^{*}NK}}{4\pi}\frac{|\vec{p}_{K(s)}|^3(2\sqrt{s}
|\vec{p}_{K(s)}|^2+m^{2}_{K}(\sqrt{m^{2}_{N}+|\vec{p}_{K(s)}|^2}-m_{p}))}{3\sqrt{s}},\\
\Gamma_{\Lambda^{*}\Sigma\pi(s)}&=&\frac{g^{2}_{\Lambda^{*}\Sigma\pi}}{4\pi}
\frac{|\vec{p}_{\pi(s)}|^3(2\sqrt{s}|\vec{p}_{\pi(s)}|^2+m^{2}_{\pi}
(\sqrt{m^{2}_{\Sigma}+|\vec{p}_{\pi(s)}|^2}-m_{\Sigma}))}{3\sqrt{s}},\\
\Gamma_{\Lambda^{*}\Lambda\pi\pi(s)}&=&\Gamma_{\Lambda^{*}\to\Sigma^{*}_{3/2}\pi
\to\Lambda\pi\pi(s)}\times
R_{3/2}+\Gamma_{\Lambda^{*}\to\Sigma^{*}_{1/2}\pi
\to\Lambda\pi\pi(s)}\times (1-R_{3/2}), \\
\Gamma_{0}&=&0.4MeV \;\;\;for\;\;
\Gamma_{\Lambda^{*}_{1520}(\sqrt{s}=1.5196GeV)}=15.6MeV
\end{eqnarray}
  and
where
\begin{eqnarray}
|\vec{p}_{K(s)}|
&=&\frac{\sqrt{s}}{2}\sqrt{(1-\frac{(m_{K}+m_{N})^2}{s})(1-\frac{(m_{K}-m_{N})^2}{s})}\;,\\
|\vec{p}_{\pi(s)}|&=&\frac{\sqrt{s}}{2}\sqrt{(1-\frac{(m_{\pi}+m_{\Sigma})^2}{s})
(1-\frac{(m_{\pi}-m_{\Sigma})^2}{s})}\;
\end{eqnarray}
are the magnitudes of the three momenta of the $K$ and $\pi$ mesons;
\begin{eqnarray}
\Gamma_{\Lambda^{*}\to\Sigma^{*}_{3/2}\pi \to\Lambda\pi\pi(s)}&=&
\int|M_{\Lambda^{*}\to\Sigma^{*}_{3/2}\pi \to\Lambda\pi\pi(s)}
B1_{(Q_{\Sigma^{*}_{3/2}\Lambda\pi})}F_{\Sigma^{*}_{3/2}}|^{2}d\phi_{\Lambda^{*}\to\Lambda\pi\pi},\\
\Gamma_{\Lambda^{*}\to\Sigma^{*}_{3/2}\pi
\to\Lambda\pi\pi(s)}&=&\int|M_{\Lambda^{*}\to\Sigma^{*}_{1/2}\pi
\to\Lambda\pi\pi(s)}B1_{(Q_{\Lambda^{*}\Sigma^{*}_{1/2}\pi})}F_{\Sigma^{*}_{1/2}}|^{2}
d\phi_{\Lambda^{*}\to\Lambda\pi\pi},
\label{bp4}
\end{eqnarray}
are the decay widths for the processes
$\Lambda^{*}\to\Sigma^{*}_{3/2}\pi \to\Lambda\pi\pi$ and
$\Lambda^{*}\to\Sigma^{*}_{1/2}\pi \to\Lambda\pi\pi$, respectively.
Here $M_{\Lambda^{*}\to\Sigma^{*}_{3/2}\pi \to\Lambda\pi\pi(s)}$ and
$M_{\Lambda^{*}\to\Sigma^{*}_{1/2}\pi \to\Lambda\pi\pi(s)}$ are the
corresponding amplitudes; $\phi_{\Lambda^{*}\to\Lambda\pi\pi}$ is
the phase space of $\Lambda^{*}$ decays into $\Lambda\pi\pi$; the
form factor $F_{R}$ and Blatt-Weisskopf centrifugal barrier factor
$B1_{(Q_{abc})}$ are given in Eqs.(\ref{ff1}) and (\ref{bl}),
respectively; the parameter $R_{3/2}$ stands for the proportion of
$\Sigma^{*}_{3/2}$ in the $\Sigma^{*}$.

Since the baryons and mesons are not point-like particles we need to
consider the form factors for each interaction vertices in order to
calculate amplitudes for the reaction. Therefore now we give the
form factors for every Feynman diagram. For the Fig.\ref{fe}(a),  we
use the following form factors
\begin{eqnarray}\label{ff1}
F_R(q^2)=\frac{\Lambda^{4}}{\Lambda^{4}+(q^{2}-m^{2}_{R})^{2}}
\end{eqnarray}
with $\Lambda=0.8GeV$ ( $R=\Lambda^{*}$ or $\Sigma^{*}$ )
\cite{ouyang}. Because both the $\Lambda^{*}$ and $\Sigma^{*}$ are
almost on-shell, the contribution of these form factors are
unimportant. In addition, we also use the following P-wave and
D-wave Blatt-Weisskopf barrier form factors for the vertices of
$\Lambda^{*}\Sigma^{*}_{1/2}\pi$, $\Sigma^{*}_{3/2}\Lambda \pi$ and
$\Lambda^{*}NK$
\begin{eqnarray}\label{bl}
B1_{(Q_{abc})}&=&\sqrt{\frac{\tilde{Q}^{2}_{abc}+
Q^{2}_{0}}{Q^{2}_{abc} + Q^{2}_{0}}},\\
B2_{(Q_{abc})}&=&\sqrt{\frac{\tilde{Q}^{4}_{abc}+3\tilde{Q}^{2}_{abc}Q^{2}_{0}+
9Q^{4}_{0}}{Q^{4}_{abc}+3Q^{2}_{abc}Q^{2}_{0}+9Q^{4}_{0}}},
\end{eqnarray}
with
\begin{eqnarray}
Q^{2}_{abc}&=&\frac{(s_{a}+s_{b}-s_{c})^{2}}{4s_{a}}-s_{b},\\
\tilde{Q}^{2}_{abc}&=&\frac{(m^{2}_{a}+m^{2}_{b}-m^{2}_{c})^{2}}{4m^{2}_{a}}-m^{2}_{b}.
\end{eqnarray}
Here $Q_{0} = 0.197321/R$ is a hadron scale parameter in the unit
of GeV/c with R the radius of the centrifugal barrier in the unit
of fm. In our calculation we set $R=0.2fm$. We find that these two
form factors have negligible effect on our results, thus one may
conclude that Fig.\ref{fe}(a) is almost model independent.

For the Figs.\ref{fe}(b,c), we  use Eq.(\ref{bl}) for the
$\Sigma^{*}_{3/2}\Lambda \pi$ vertex, and Eq.(\ref{ff1}) for the
off-shell baryon resonance $\Sigma^{*}_{3/2}$ with cut-off
parameter $\Lambda=1.0$ GeV for $K^{*}$ exchange and $n$ exchange.

For the Figs.\ref{fe}(d,e,f), we also use the  form factor in
Eq.(\ref{ff1}) with  $\Lambda=1.0$ GeV for  $K^{*}$ and $K^{-}$
exchange diagram, and $\Lambda=1.8$ GeV for $n$ and $p$ exchange
diagram.

After fixing the relevant effective Lagrangians, coupling constants,
propagators and form factors, the amplitudes for various Feynman
diagrams can be written down straightforwardly by following the
Feynman rules, and total amplitude is just their simple sum.
Here as an example, we give explicitly the individual amplitudes
corresponding to $\Lambda^{*}\to\Sigma^{*}_{3/2}\pi$ and to
$\Lambda^{*}\to\Sigma^{*}_{1/2}\pi$ for the Feynman diagrams (a) in
the Fig.\ref{fe},
\begin{eqnarray}
{\cal M}_{\Lambda^{*}\to\Sigma^{*}_{3/2}\pi}&=&{\cal
M}_{\Lambda^{*}\to\Sigma^{*+}_{3/2}\pi^{-}}+{\cal
M}_{\Lambda^{*}\to\Sigma^{*-}_{3/2}\pi^{+}}\nonumber\\
&=&\!\frac{g_{\Lambda^{*}KN}g_{\Lambda^{*}\Sigma^{*}_{3/2}\pi}g_{\Sigma^{*}_{3/2}\Lambda\pi}}{\sqrt{6}}
F_{\Lambda^{*}}B2_{(Q_{\Lambda^{*}NK})}\bar{u}_{p_{\Lambda}s_{\Lambda}}
\large(p^{\alpha}_{\pi^{+}}G^{(\frac{3}{2})}_{\Sigma^{*+}_{3/2}\alpha\mu}F_{\Sigma^{*+}_{3/2}}
B1_{(Q_{\Sigma^{*+}_{3/2}\Lambda\pi^{+}})}\nonumber\\
&&+p^{\alpha}_{\pi^{-}}G^{(\frac{3}{2})}_{\Sigma^{*-}_{3/2}\alpha\mu}F_{\Sigma^{*-}_{3/2}}
B1_{(Q_{\Sigma^{*-}_{3/2}\Lambda\pi^{-}})}\large)
G^{(\frac{3}{2})\mu\nu}_{\Lambda^{*}}p_{K^{-}\nu}\!\gamma_{5}\!\!\not\!p_{K^-}
u_{p_{p}s_{p}},\\
{\cal M}_{\Lambda^{*}\to\Sigma^{*}_{1/2}\pi}&=&{\cal
M}_{\Lambda^{*}\to\Sigma^{*+}_{1/2}\pi^{-}}+{\cal
M}_{\Lambda^{*}\to\Sigma^{*-}_{1/2}\pi^{+}}\nonumber\\
&=&\!\frac{g_{\Lambda^{*}KN}g_{\Lambda^{*}\Sigma^{*}_{1/2}\pi}g_{\Sigma^{*}_{1/2}\Lambda\pi}}{\sqrt{6}}
F_{\Lambda^{*}}B2_{(Q_{\Lambda^{*}NK})}\bar{u}_{p_{\Lambda}s_{\Lambda}}
(G^{(\frac{1}{2})}_{\Sigma^{*+}_{1/2}}p_{\pi^{-}\mu}F_{\Sigma^{*+}_{1/2}}
B1_{(Q_{\Lambda^{*}\Sigma^{*+}_{1/2}\pi^{-}})}\nonumber\\
&&
+G^{(\frac{1}{2})}_{\Sigma^{*-}_{1/2}}p_{\pi^{+}\mu}F_{\Sigma^{*-}_{1/2}}
B1_{(Q_{\Lambda^{*}\Sigma^{*-}_{1/2}\pi^{+}})})
G^{(\frac{3}{2})\mu\nu}_{\Lambda^{*}}p_{K^{-}\nu}\!\gamma_{5}\!\!\not\!p_{K^-}
u_{p_{p}s_{p}},
\label{M1520}
\end{eqnarray}
where $u_{p_{\Lambda}s_{\Lambda}}$ and $u_{p_{p}s_{p}}$ are the spin
wave functions of the outgoing $\Lambda$  and incoming proton,
respectively; $p_{\pi^+}$, $p_{\pi^-}$ and $p_{K^{-}}$ are the 4-
momenta of the final state pions and initial state $K^{-}$ meson;
the factor $1/\sqrt{6}$ is a isospin C-G coefficient. So the total
amplitude squared for the \kp\ reaction is
\begin{eqnarray}
|{\cal M}_{K^{-}p \to \Lambda \pi^{+}\pi^{-}}|^{2}&=&|{\cal
M}_{\Lambda^{*}\to\Sigma^{*}_{3/2}\pi}|^{2}\times R_{3/2}+|{\cal
M}_{\Lambda^{*}\to\Sigma^{*}_{1/2}\pi}|^{2}\times (1-R_{3/2})+
\nonumber\\
&&|{\cal M}_{n\Sigma^{*-}_{3/2}\pi^{+}}|^{2}+|{\cal
M}_{K^{*0}\Sigma^{*-}_{3/2}\pi^{+}}|^{2}+\nonumber\\
&&|{\cal M}_{K^{*0}n}|^{2}+|{\cal M}_{K^{*0}K^{-}}|^{2}+|{\cal
M}_{pn}|^{2}.\label{M}
\end{eqnarray}
Note that we do not include the interference terms between
different Feynman diagrams because their contributions are
insignificant. Then the calculation of the cross section for \kp\
is straightforward:
\begin{eqnarray}
d\sigma_{K^{-}p \to \Lambda \pi^{+}\pi^{-}}&=&
\frac{1}{4}\frac{m_{p}}{\sqrt{(p_{p}\cdot
p_{K^{-}})^{2}-m_{p}m_{K^{-}}}} \sum_{s_{i}}\sum_{s_{f}}|{\cal
M}_{K^{-}p \to \Lambda \pi^{+}\pi^{-}}|^{2}d\phi,\\
d\phi&=&\frac{1}{(2\pi)^{5}}\frac{m_{\Lambda}d^{3}p_{\Lambda}}{E_{\Lambda}}
\frac{d^{3}p_{\pi^{+}}}{2E_{\pi^{+}}}\frac{d^{3}p_{\pi^{-}}}{2E_{\pi^{-}}}
\delta^{4}(p_{K^{-}}+p_{p}-p_{\Lambda}-p_{\pi^{+}}-p_{\pi^{-}}).\label{corss}
\end{eqnarray}

\section{Numerical results and discussion}
\label{s3}
With the formalism and ingredients given in the former
section, we compute the total cross section versus the $K^-$ beam
momentum $Plab_{(K^{-})}$ for the \kp \ reaction for
$Plab_{(K^{-})}=0.25 - 0.60$ GeV by using the code FOWL from the CERN
program library, which is a program for Monte Carlo multi-particle
phase space integration weighted by the amplitude squared. We
consider two cases, firstly, we assume that the $J^P$ of the
$\Sigma^{*}$ is ${\frac{3}{2}}^+$ with $R_{3/2}=1.0$. On the other
hand, we suppose that the $\Sigma^{*}$ with $J^P = {\frac{3}{2}}^+$
and with $J^P = {\frac{1}{2}}^-$ account for $60\%$ and $40\%$,
respectively ( $R_{3/2}=0.60$). Our results on the total cross
section for the \kp \ reaction is almost the same for both cases,
because the branching ratio of $\Lambda^{*}(1520) \to \Lambda\pi\pi$
is $11\%$ for both cases. Total cross section, angular distributions
of the final state $\Lambda$, and  $\pi\pi$, $\Lambda\pi^+$,
$\Lambda\pi^-$ invariant mass square distributions, as well as
Dalitz plots for the final state particles for the two cases of our
theoretical calculations, are shown in Figs.2-7, with experimental
data points from the Refs.\cite{exp,exp1,exp2}.

\begin{figure}[htbp] \vspace{-0.cm}
\begin{center}
\includegraphics[width=0.9\columnwidth]{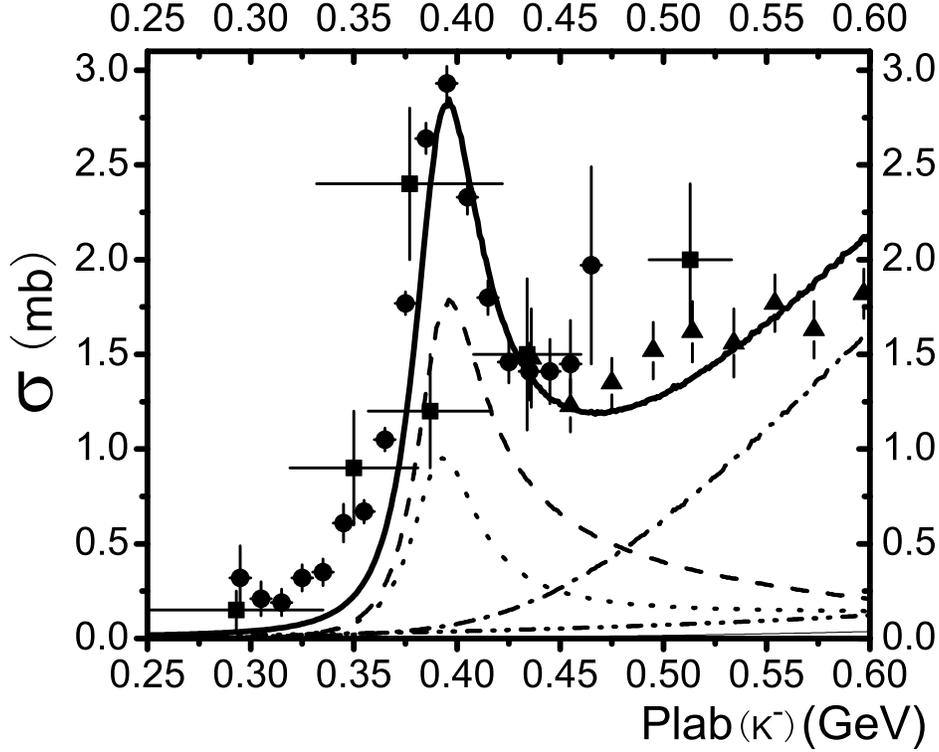}
\caption {Theoretical total cross section vs beam momentum
$Plab_{(K^{-})}$ for the $K^{-} p \to \Lambda \pi^{+} \pi^{-}$
reaction  with $R_{3/2}=0.60$. The circle, square and triangle are
data points from \cite{exp}, \cite{exp1} and \cite{exp2},
respectively. The dashed and dotted curves are for the Fig.1(a) with
$\Sigma^{*}_{3/2}$ and $\Sigma^{*}_{1/2}$, respectively; the
dash-dotted and the dash-dot-dotted curves for the Fig.1(b) and
Fig.1(d), respectively; curves close to zero for Fig.1(c,e,f). The
solid curve is the sum of these broken curves.} \label{all}
\end{center}
\end{figure}

\begin{figure}[htbp] \vspace{-0.cm}
\begin{center}
\includegraphics[width=0.32\columnwidth]{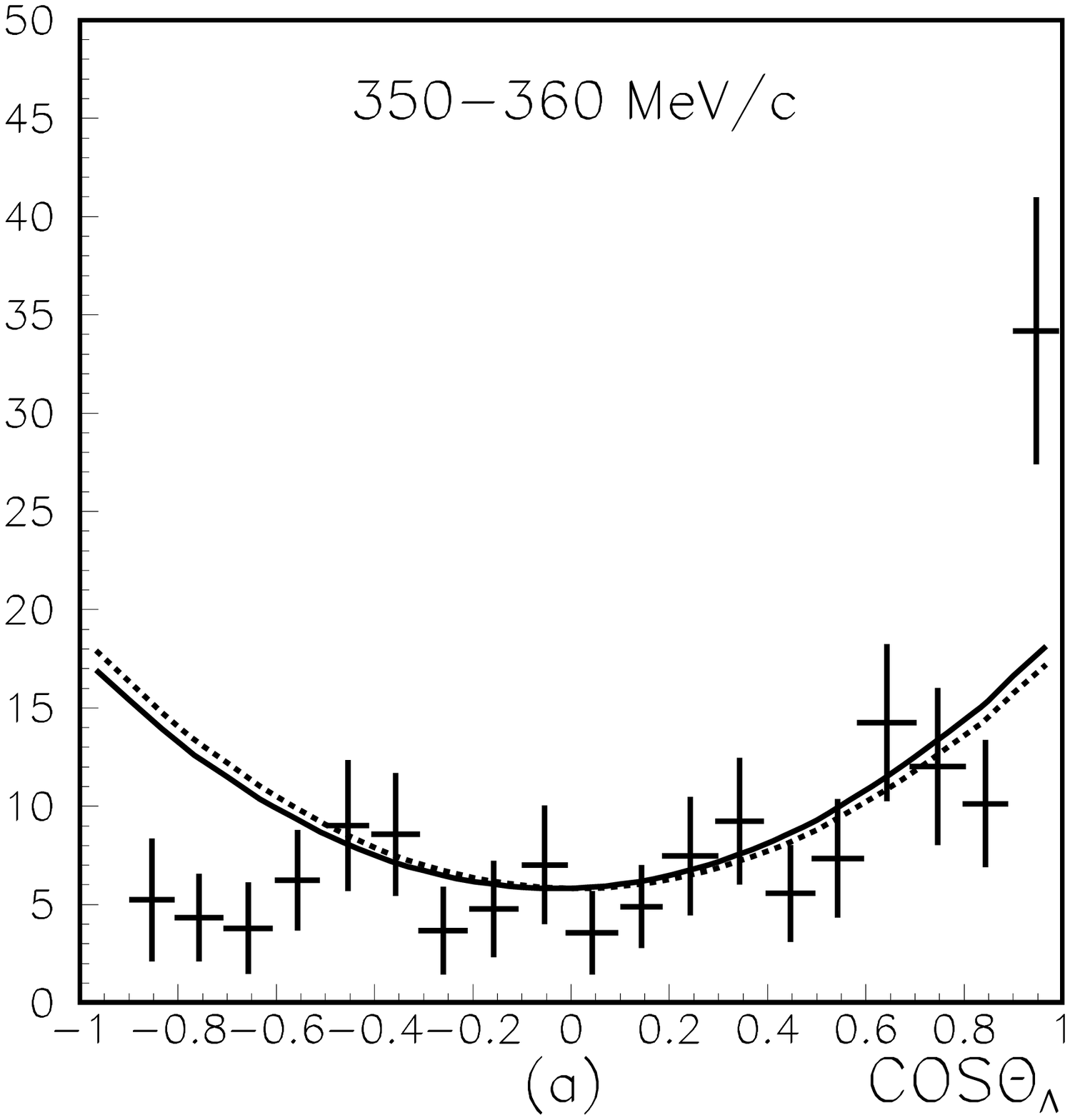}
\includegraphics[width=0.32\columnwidth]{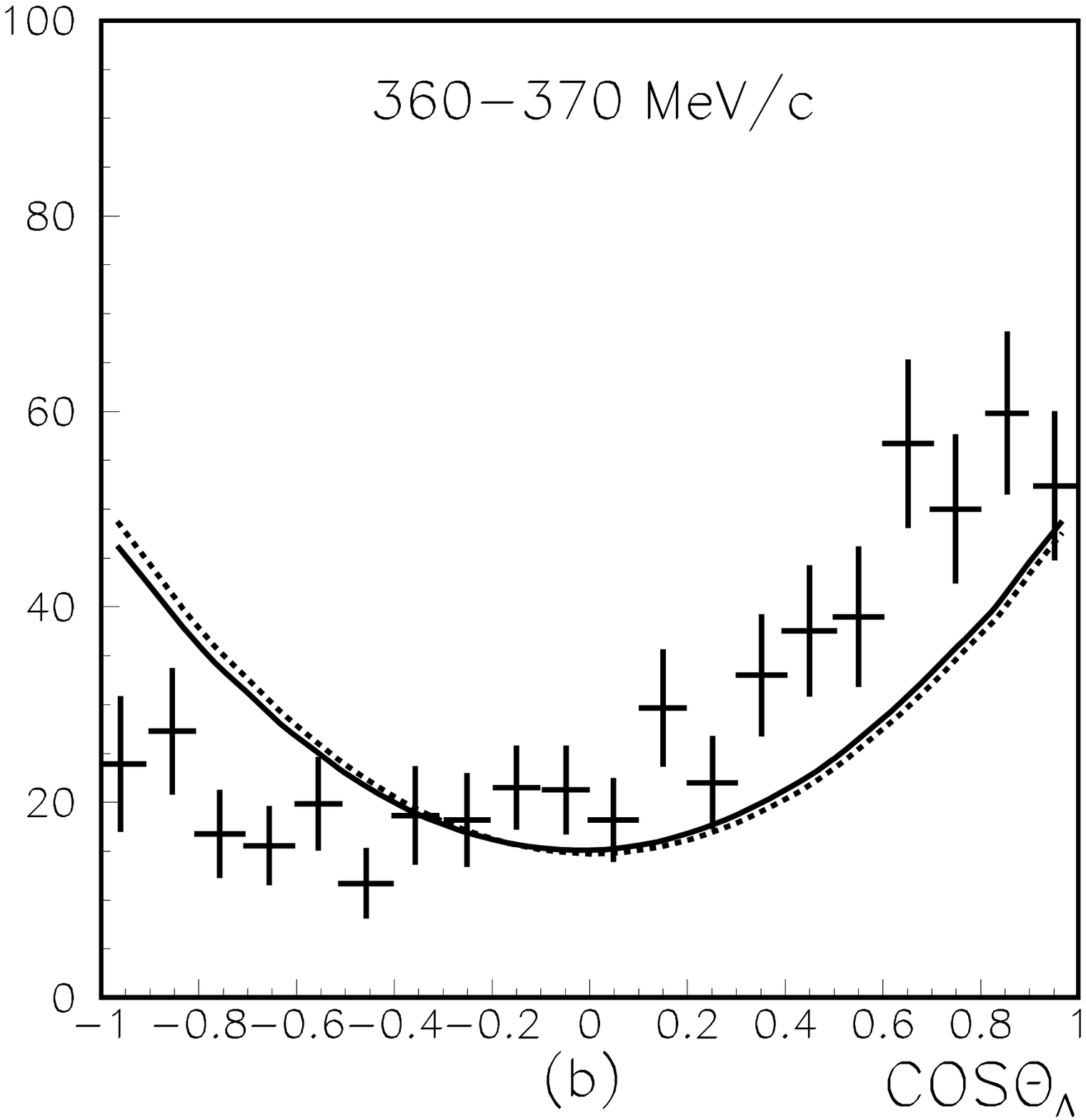}
\includegraphics[width=0.32\columnwidth]{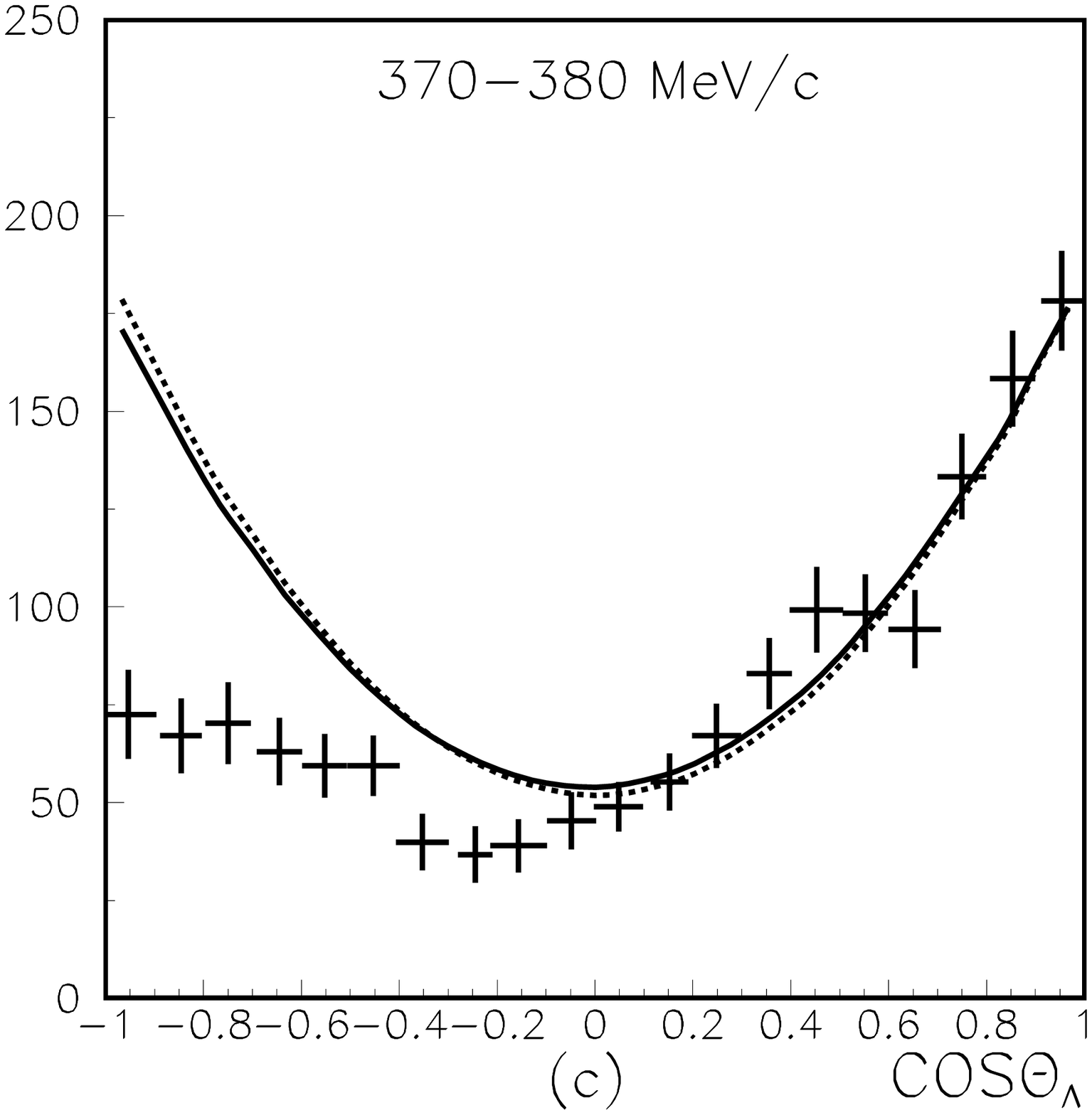}
\includegraphics[width=0.32\columnwidth]{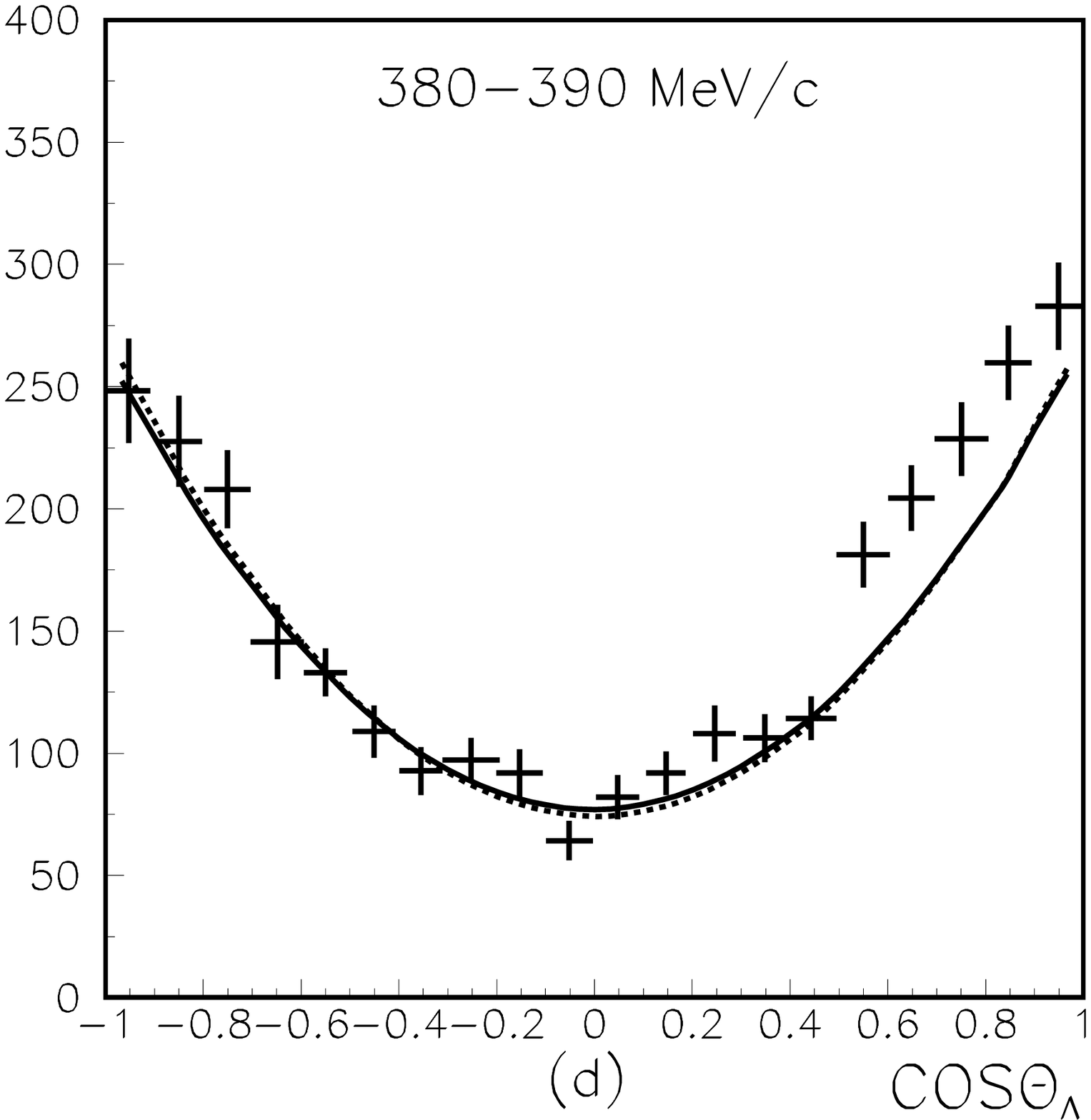}
\includegraphics[width=0.32\columnwidth]{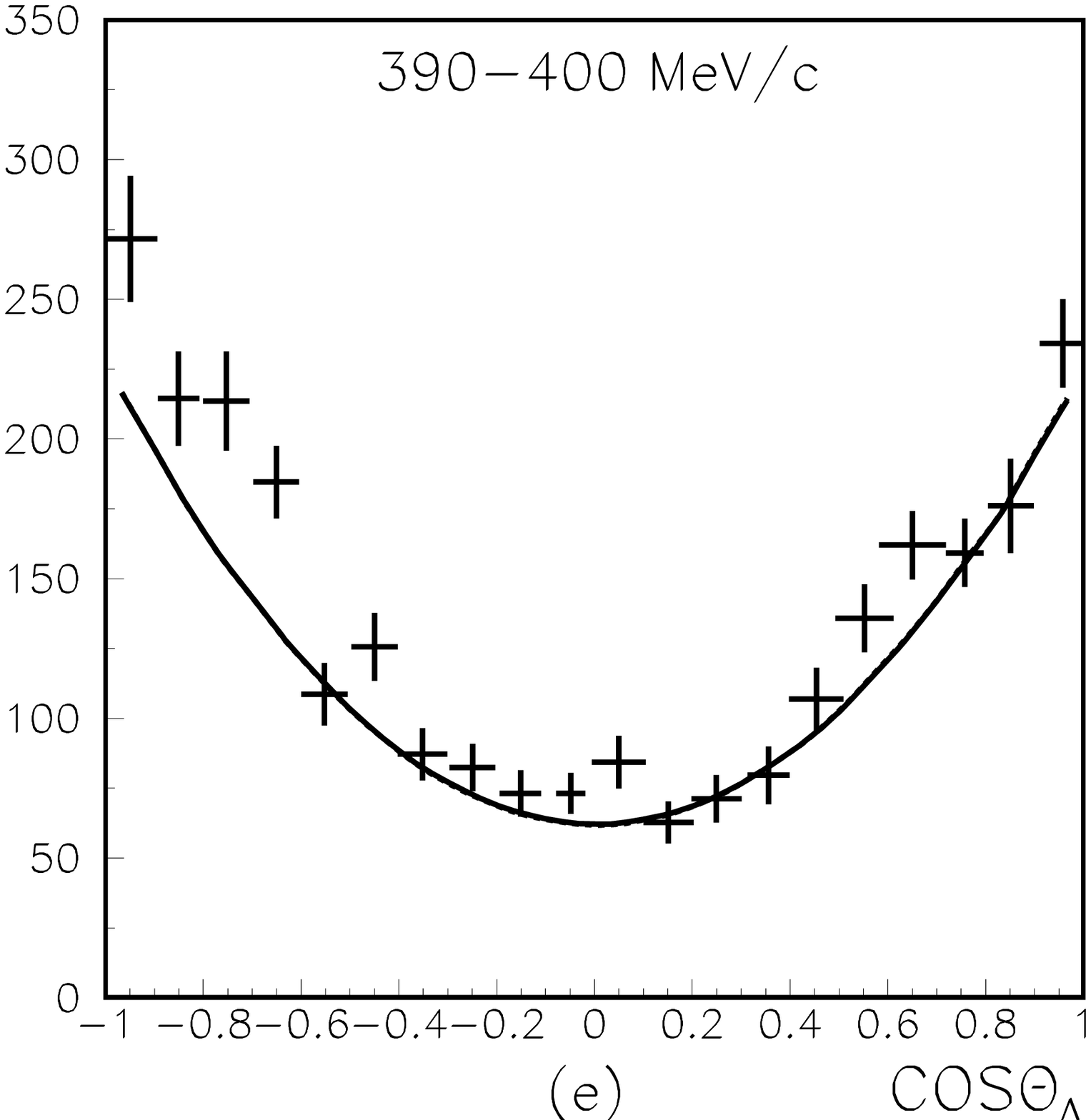}
\includegraphics[width=0.32\columnwidth]{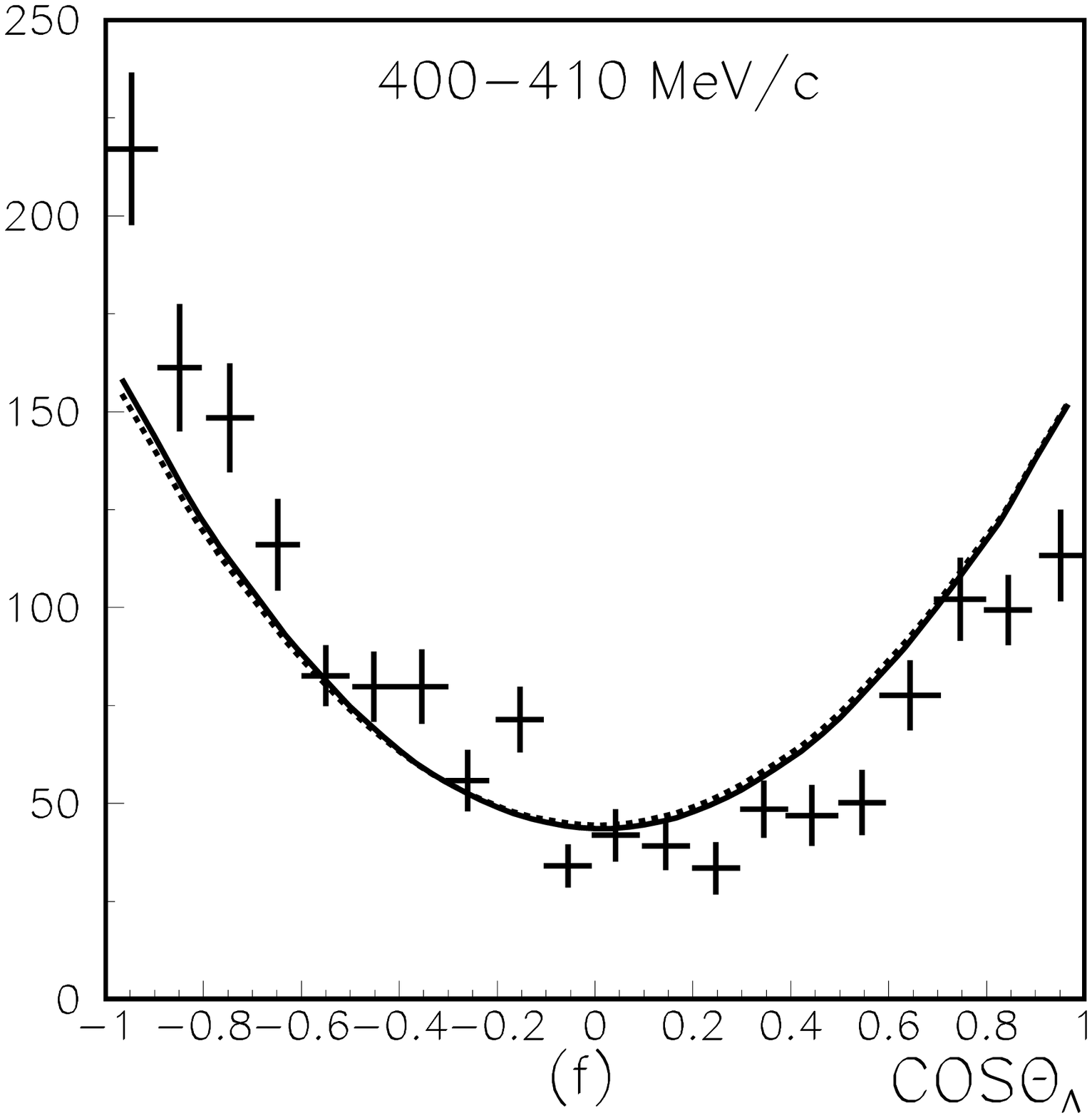}
\includegraphics[width=0.32\columnwidth]{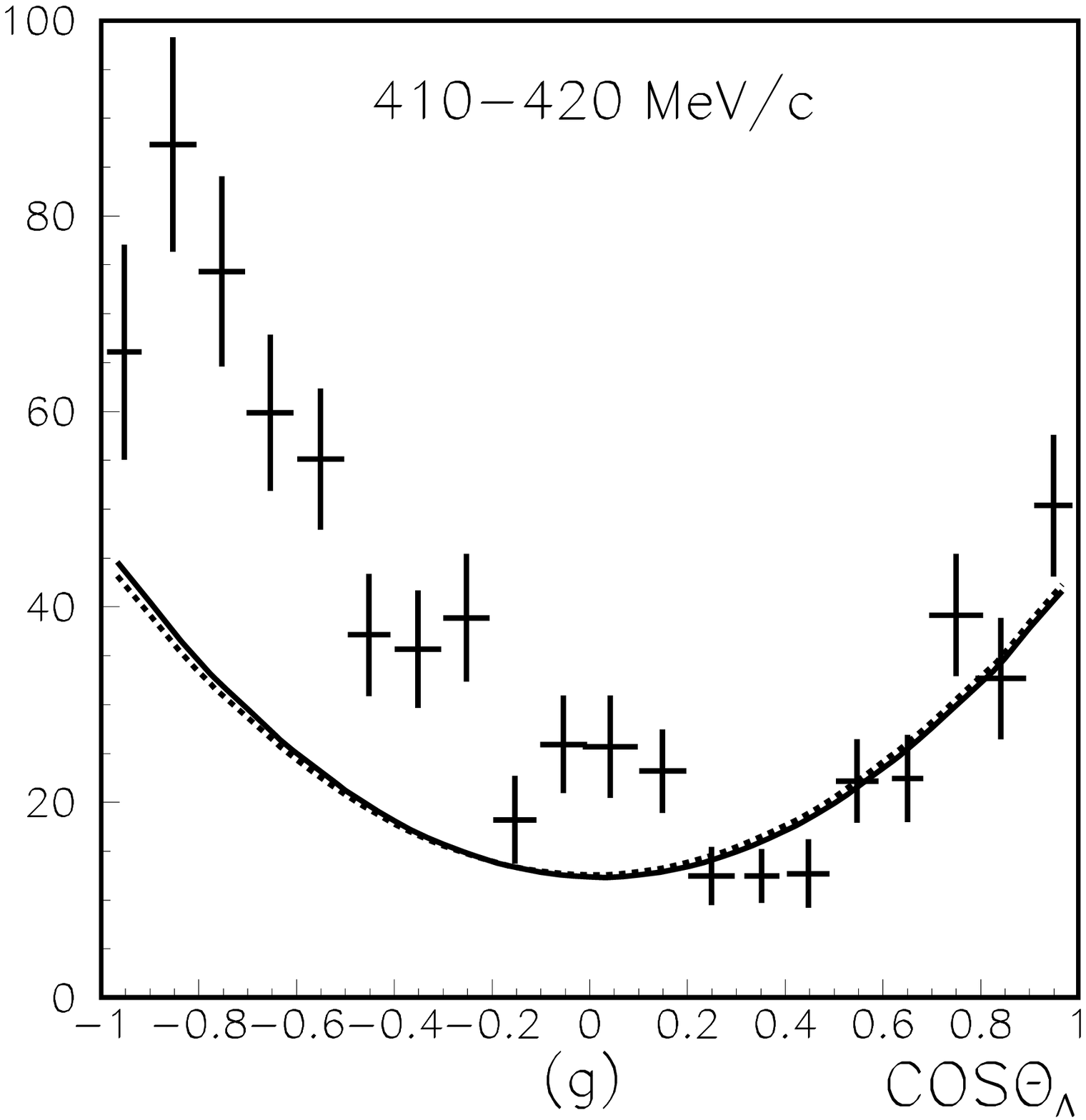}
\includegraphics[width=0.32\columnwidth]{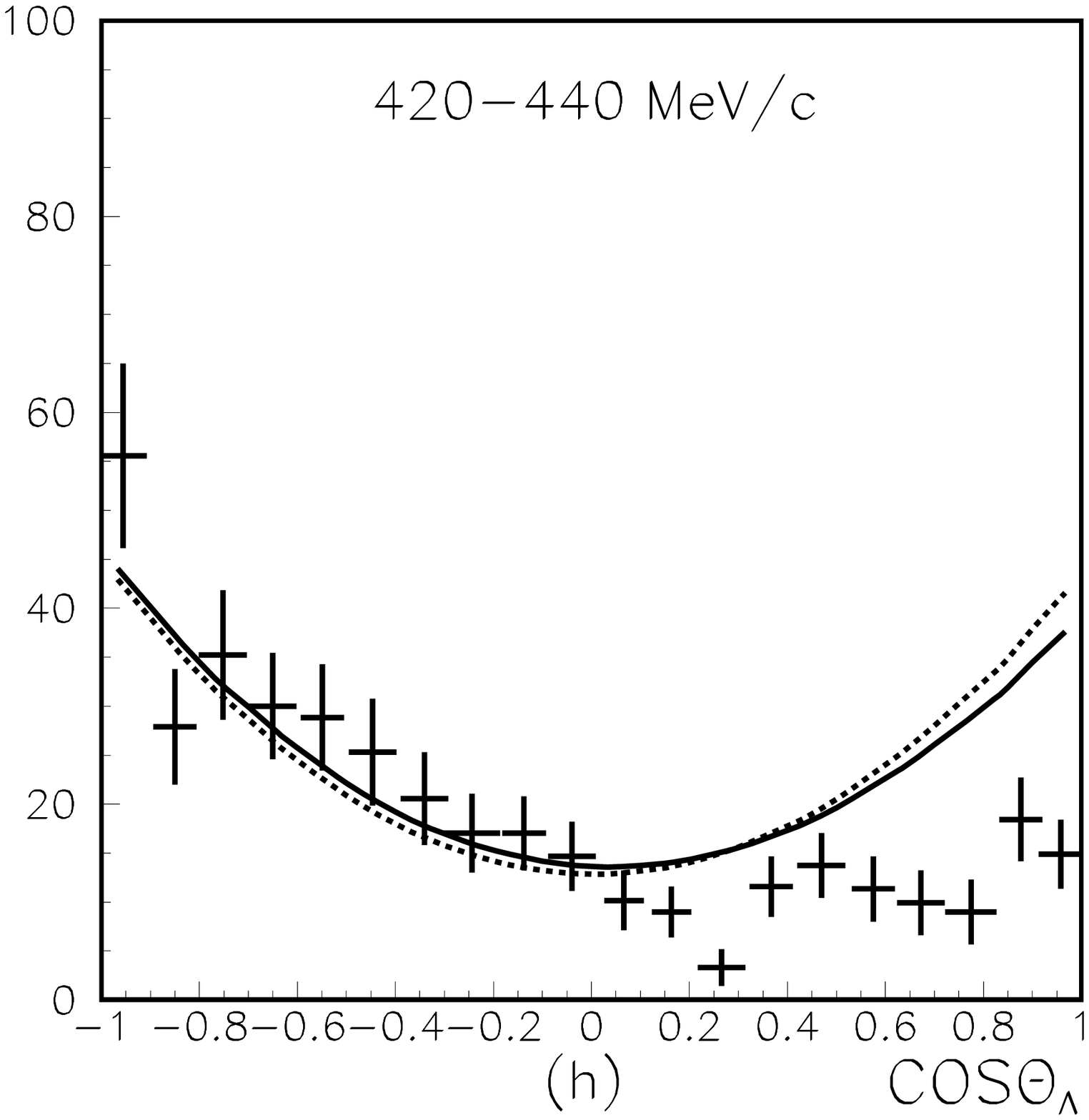}
\includegraphics[width=0.32\columnwidth]{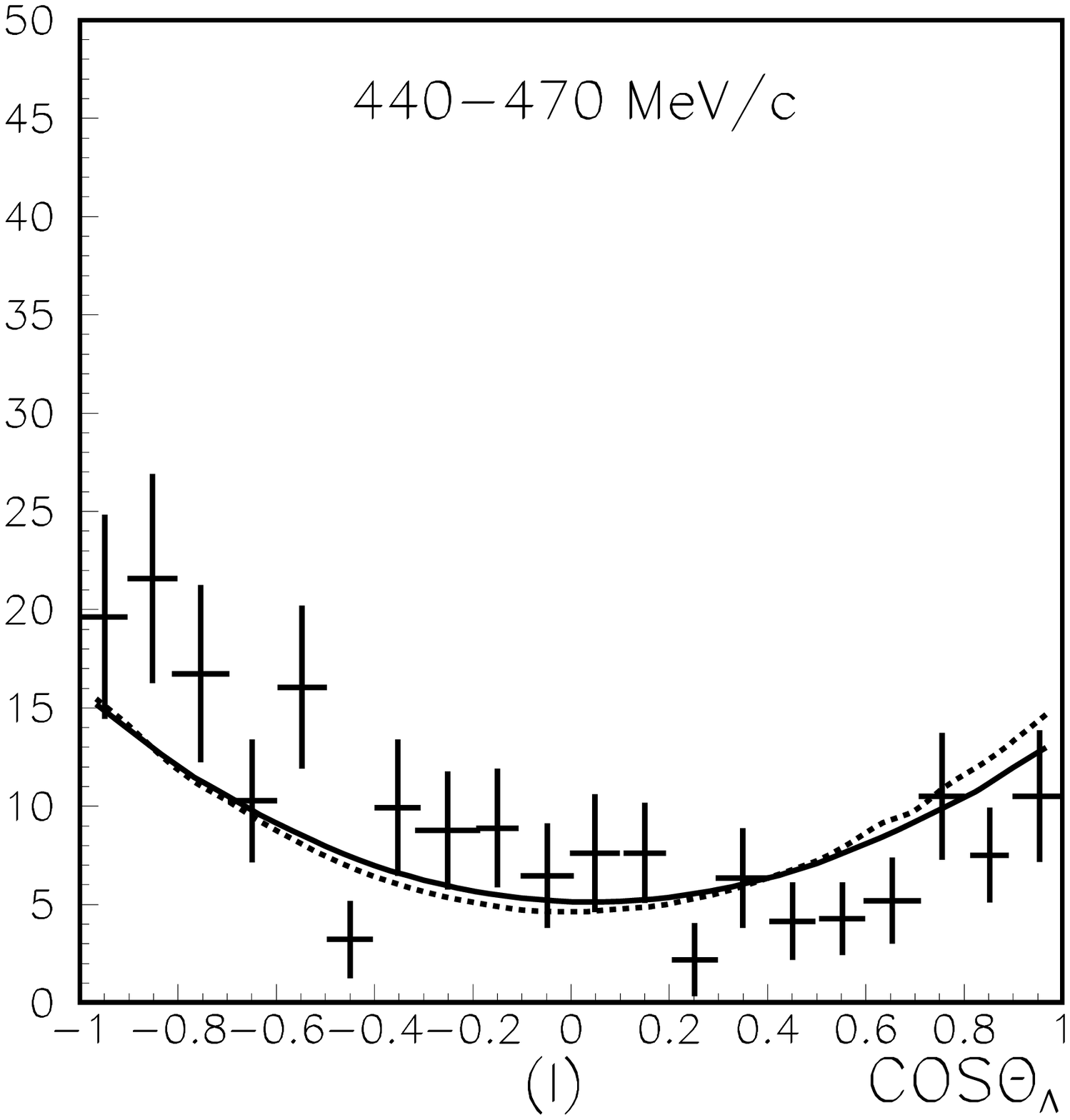}
\caption{Theoretical angular distribution of the $\Lambda$ for
various $K^-$ beam momenta compared with data~\cite{exp}. The dotted
line is for the pure $\Sigma^{*}(1385)$ with $J^P =
{\frac{3}{2}}^+$; the solid line includes $\Sigma^{*}_{1/2}$ in
addition with $R_{3/2}=0.60$.} \label{alam}
\end{center}
\end{figure}

\begin{figure}[htbp] \vspace{-0.cm}
\begin{center}
\includegraphics[width=0.32\columnwidth]{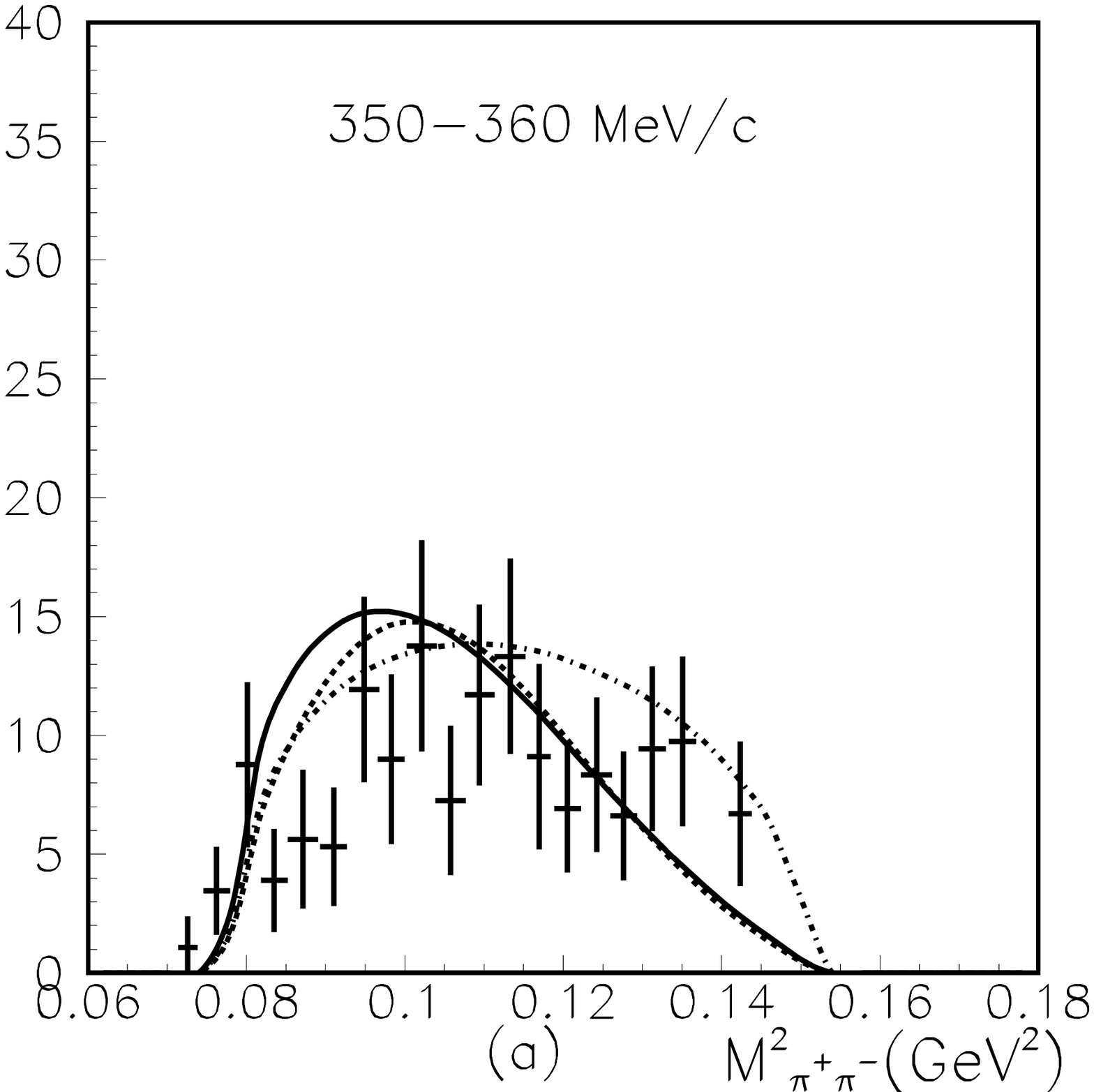}
\includegraphics[width=0.32\columnwidth]{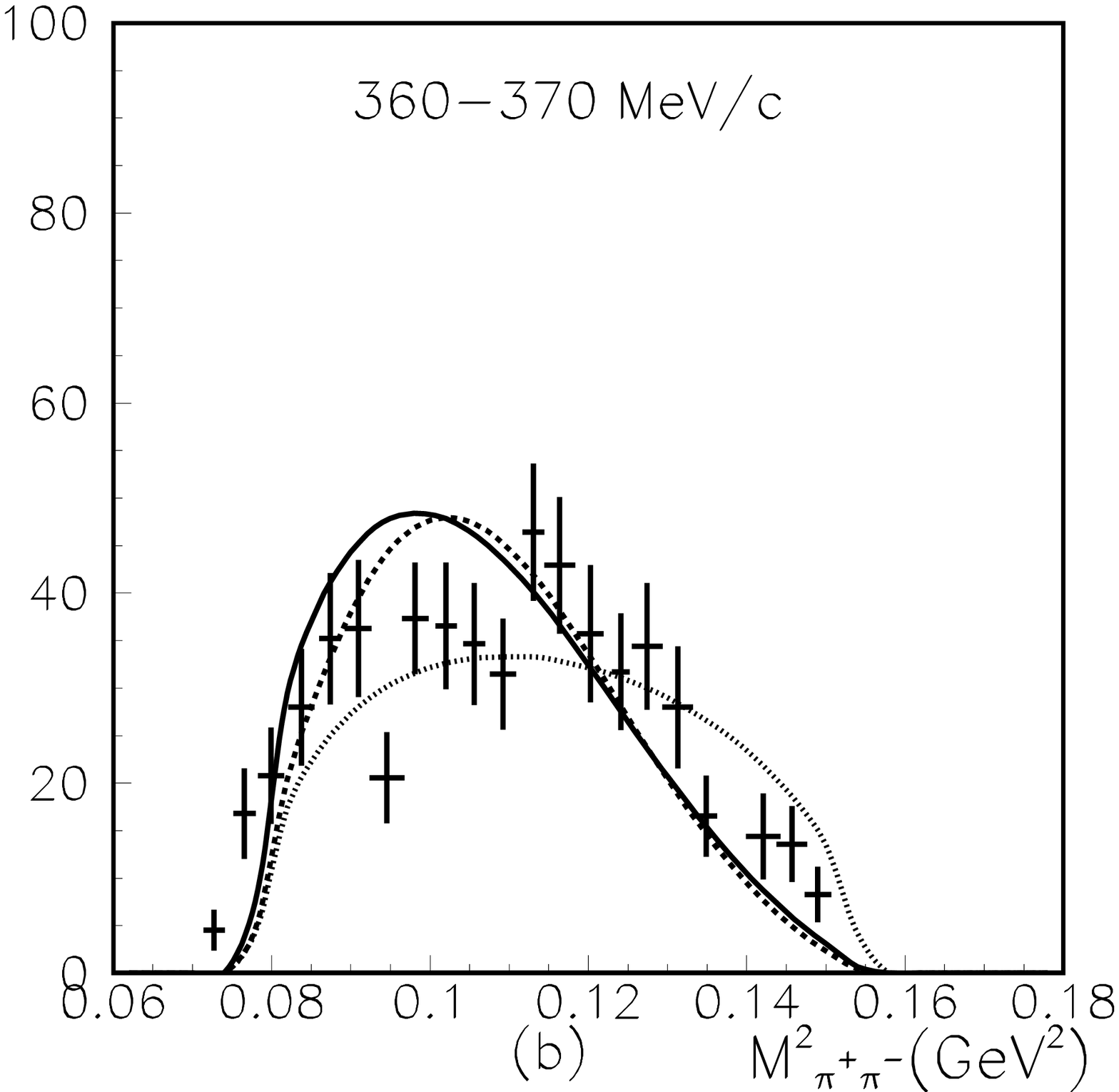}
\includegraphics[width=0.32\columnwidth]{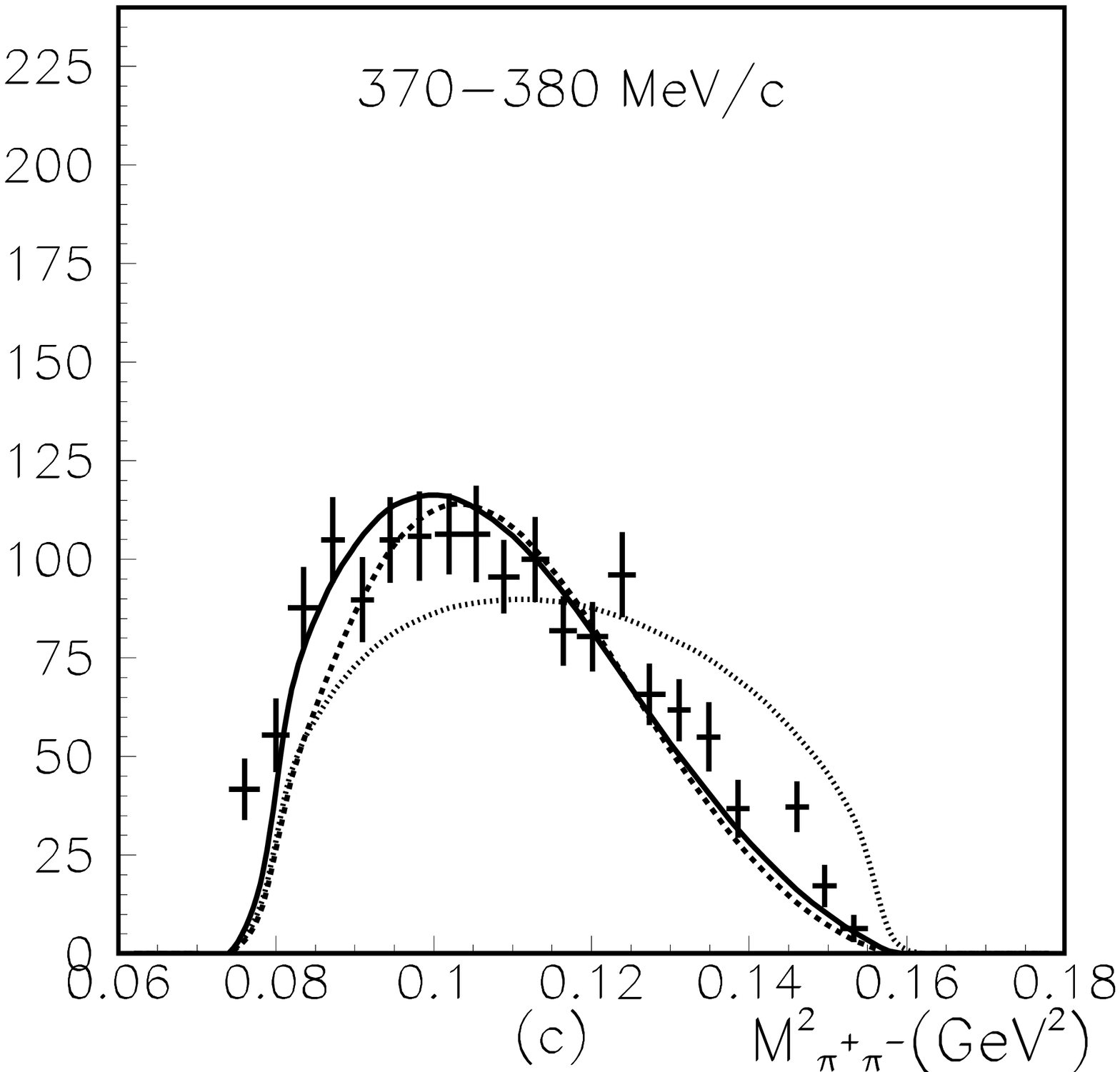}
\includegraphics[width=0.32\columnwidth]{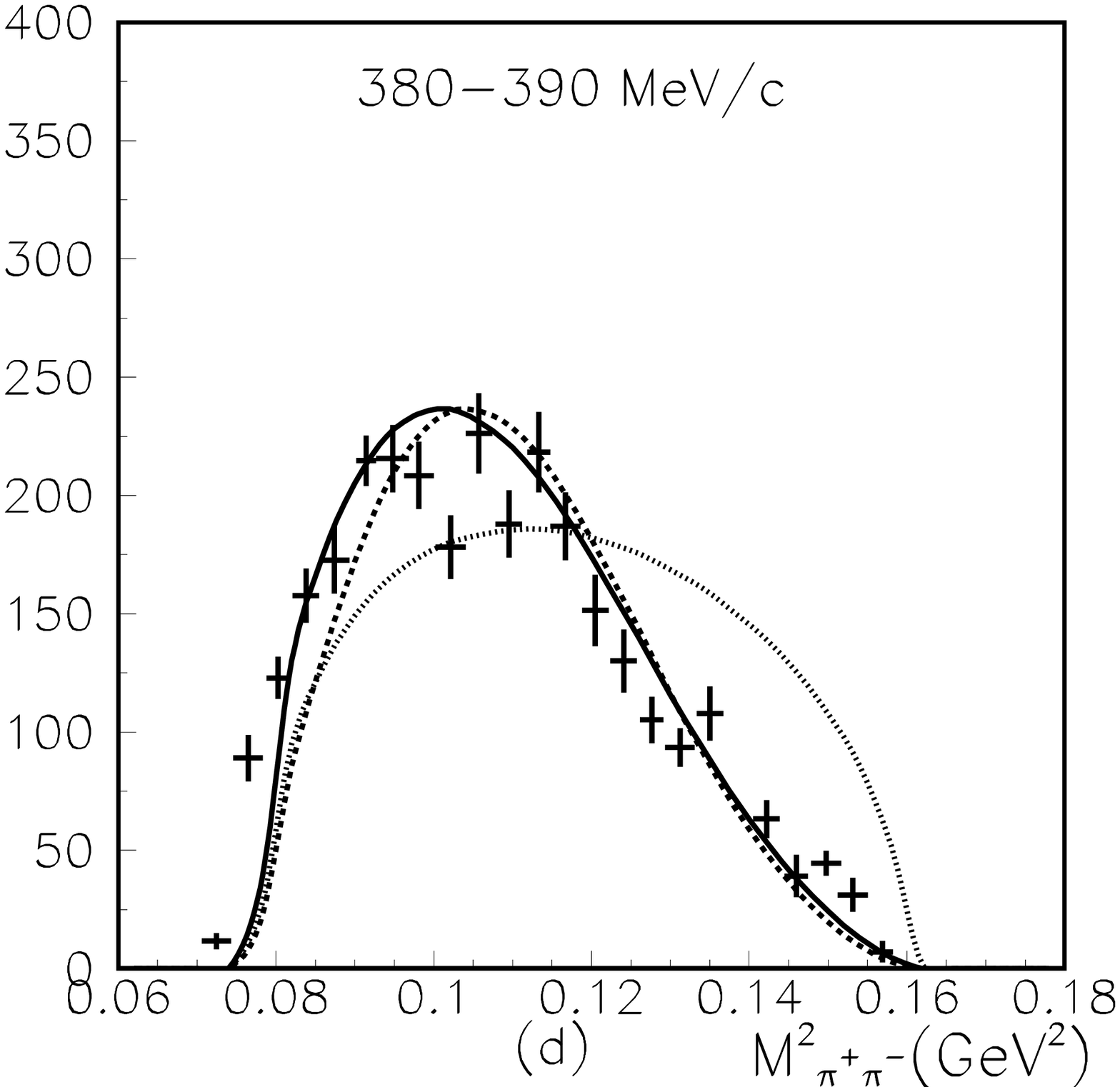}
\includegraphics[width=0.32\columnwidth]{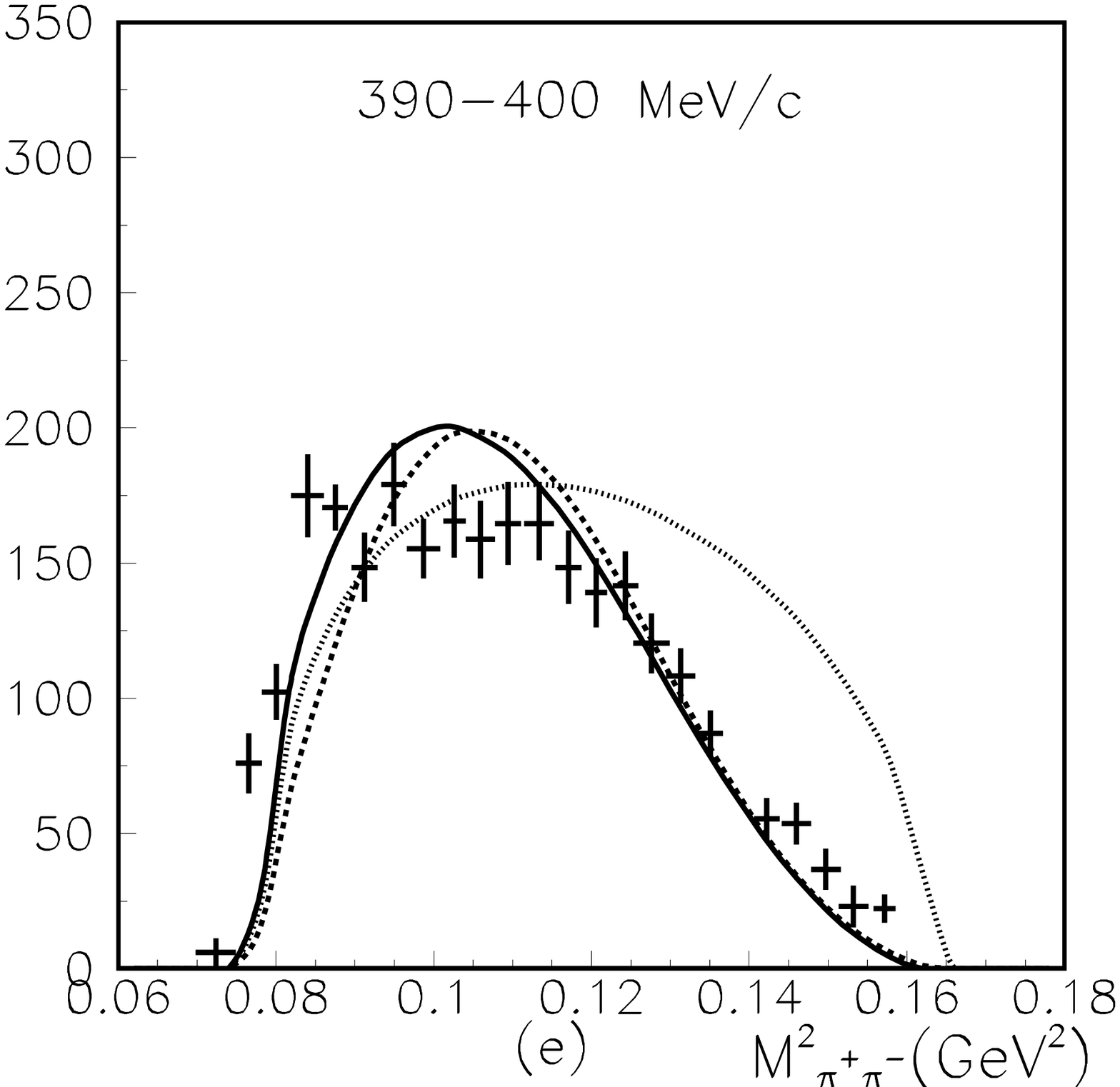}
\includegraphics[width=0.32\columnwidth]{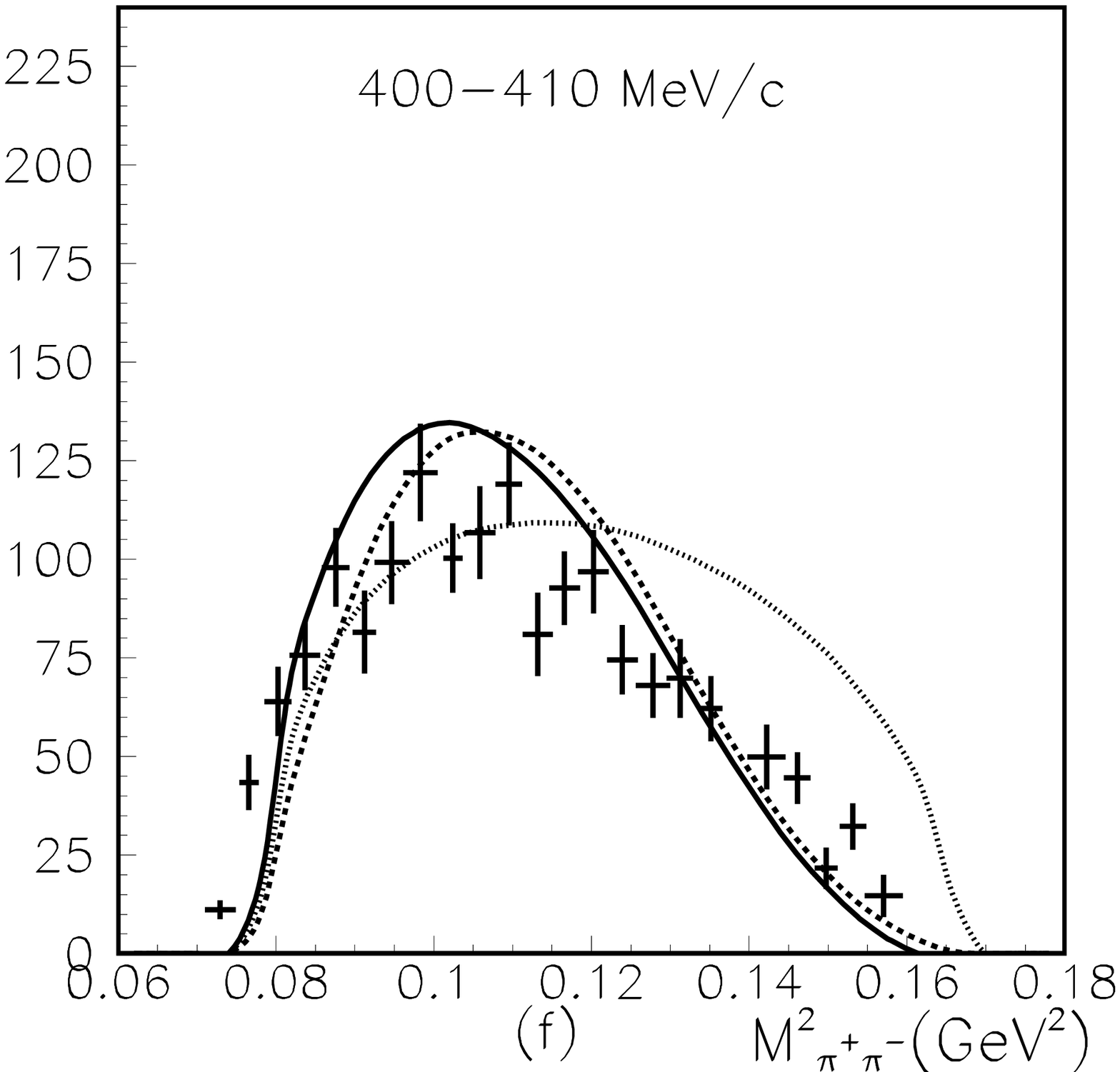}
\includegraphics[width=0.32\columnwidth]{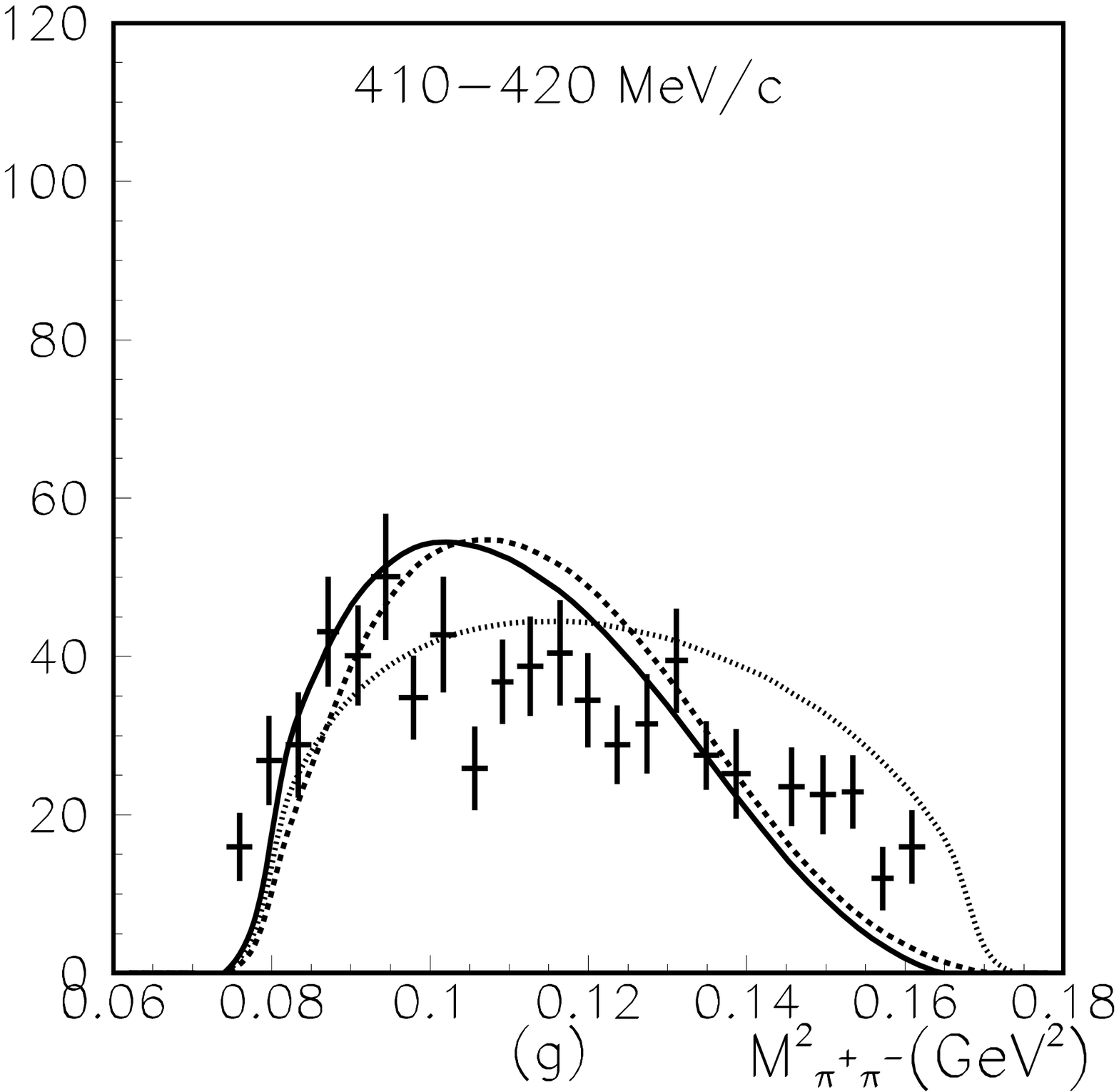}
\includegraphics[width=0.32\columnwidth]{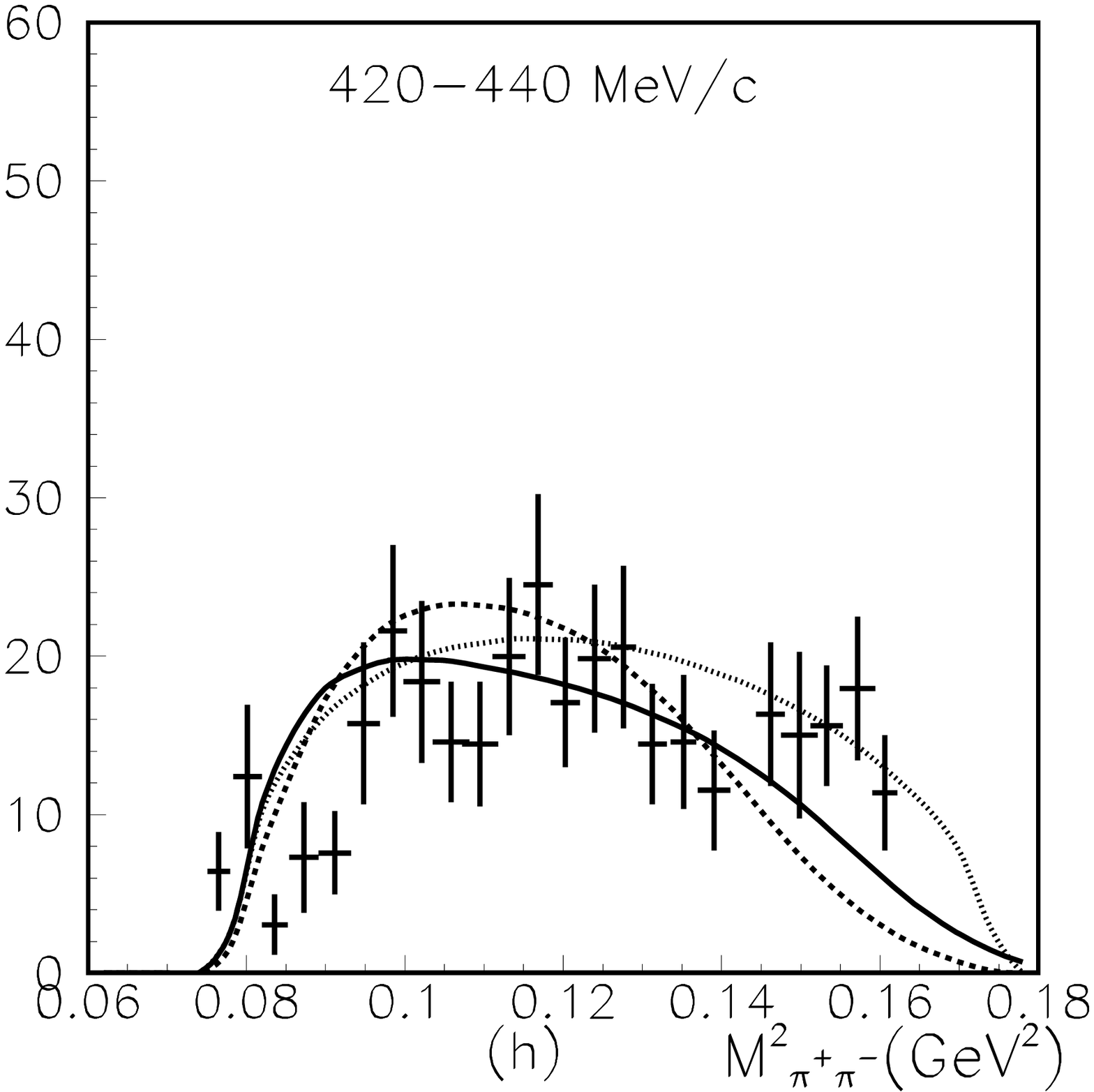}
\includegraphics[width=0.32\columnwidth]{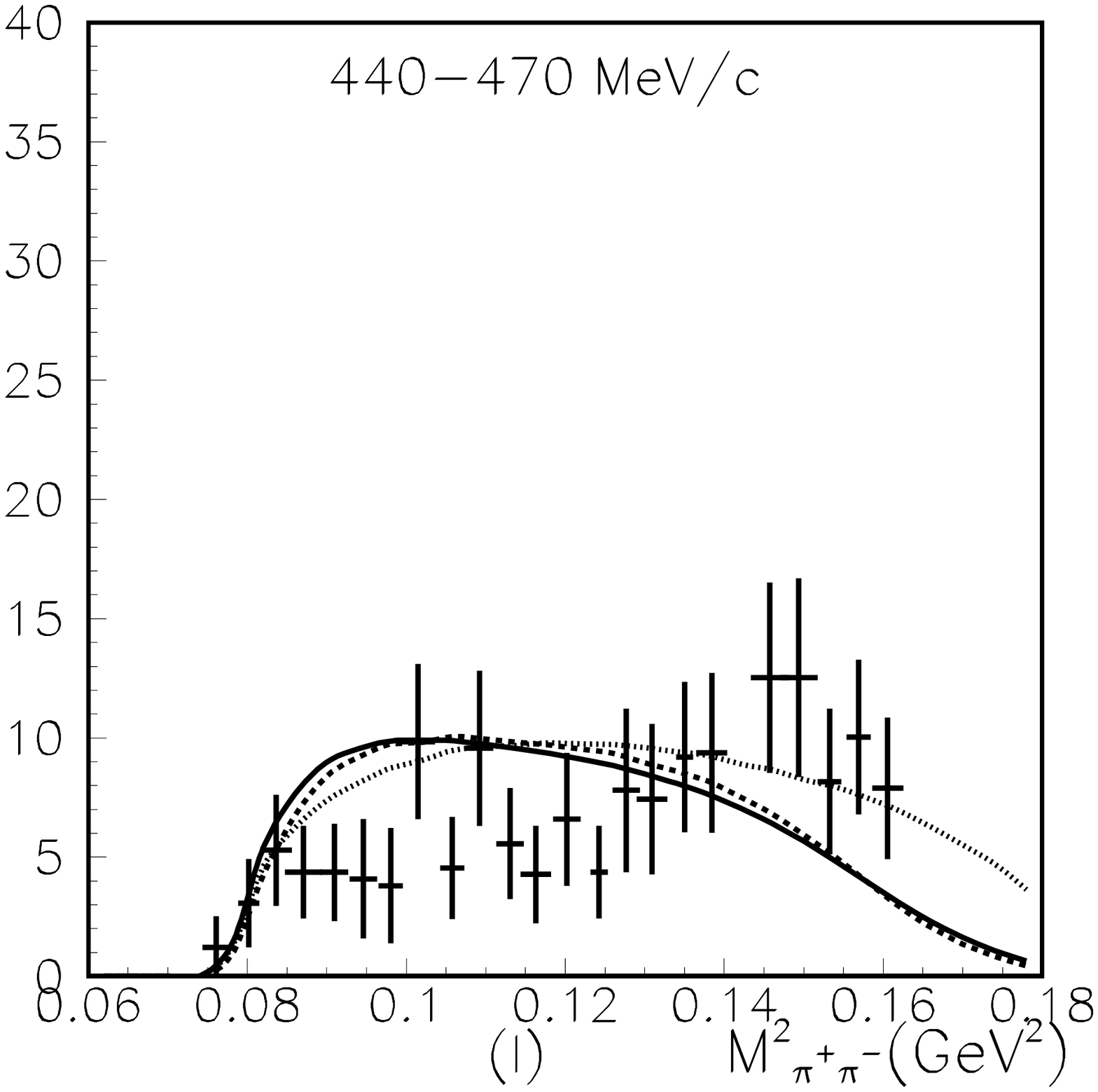}
\caption{Theoretical $\pi^+ \pi^-$ invariant mass  squared
distribution for various $K^-$ beam momenta compared with
data~\cite{exp}. The dotted line is for the pure $\Sigma^{*}(1385)$
with $J^P = {\frac{3}{2}}^+$; the solid line includes
$\Sigma^{*}_{1/2}$ in addition with $R_{3/2}=0.60$.} \label{pipi}
\end{center}
\end{figure}

\begin{figure}[htbp] \vspace{-0.cm}
\begin{center}
\includegraphics[width=0.3\columnwidth]{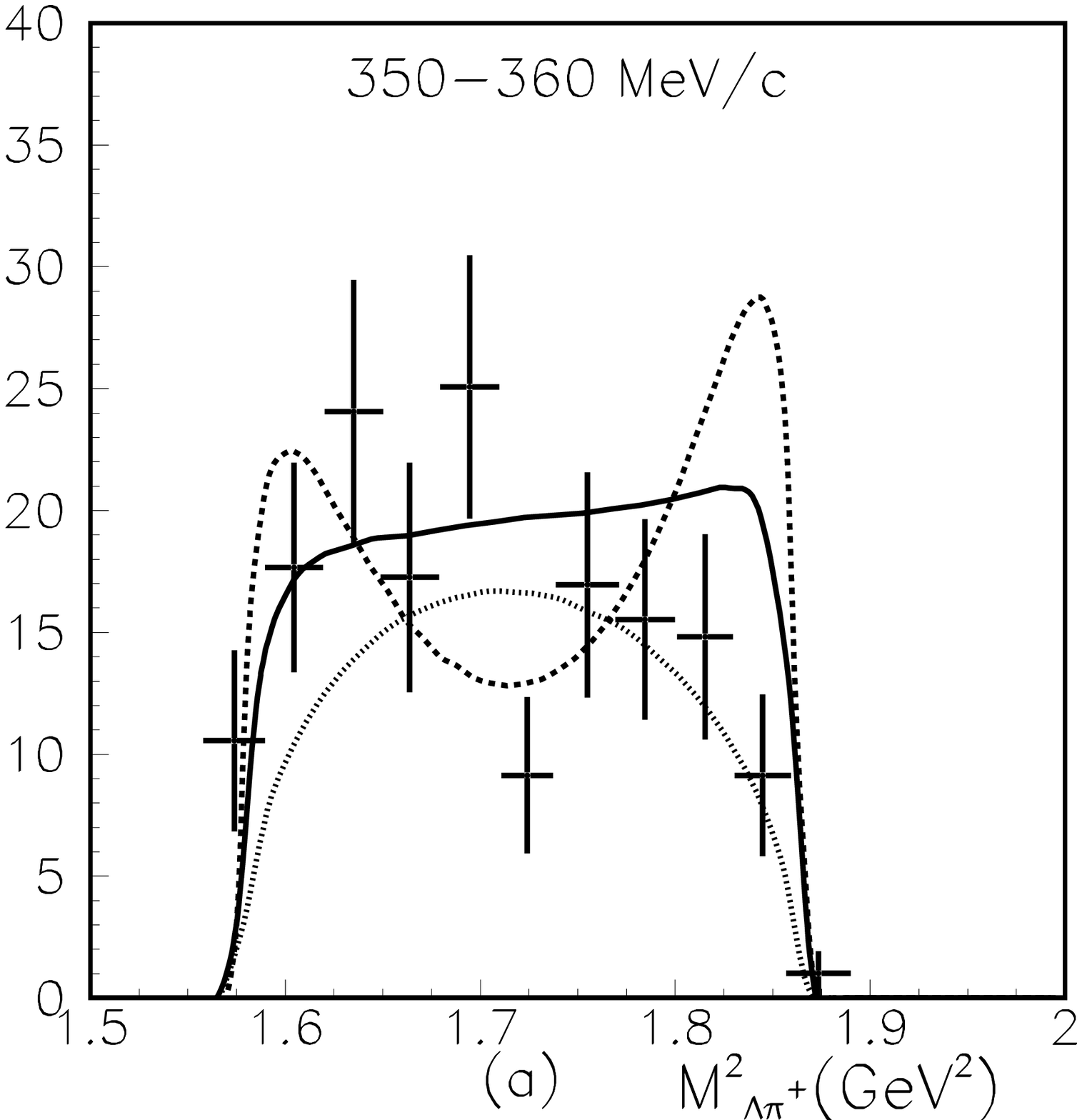}
\includegraphics[width=0.3\columnwidth]{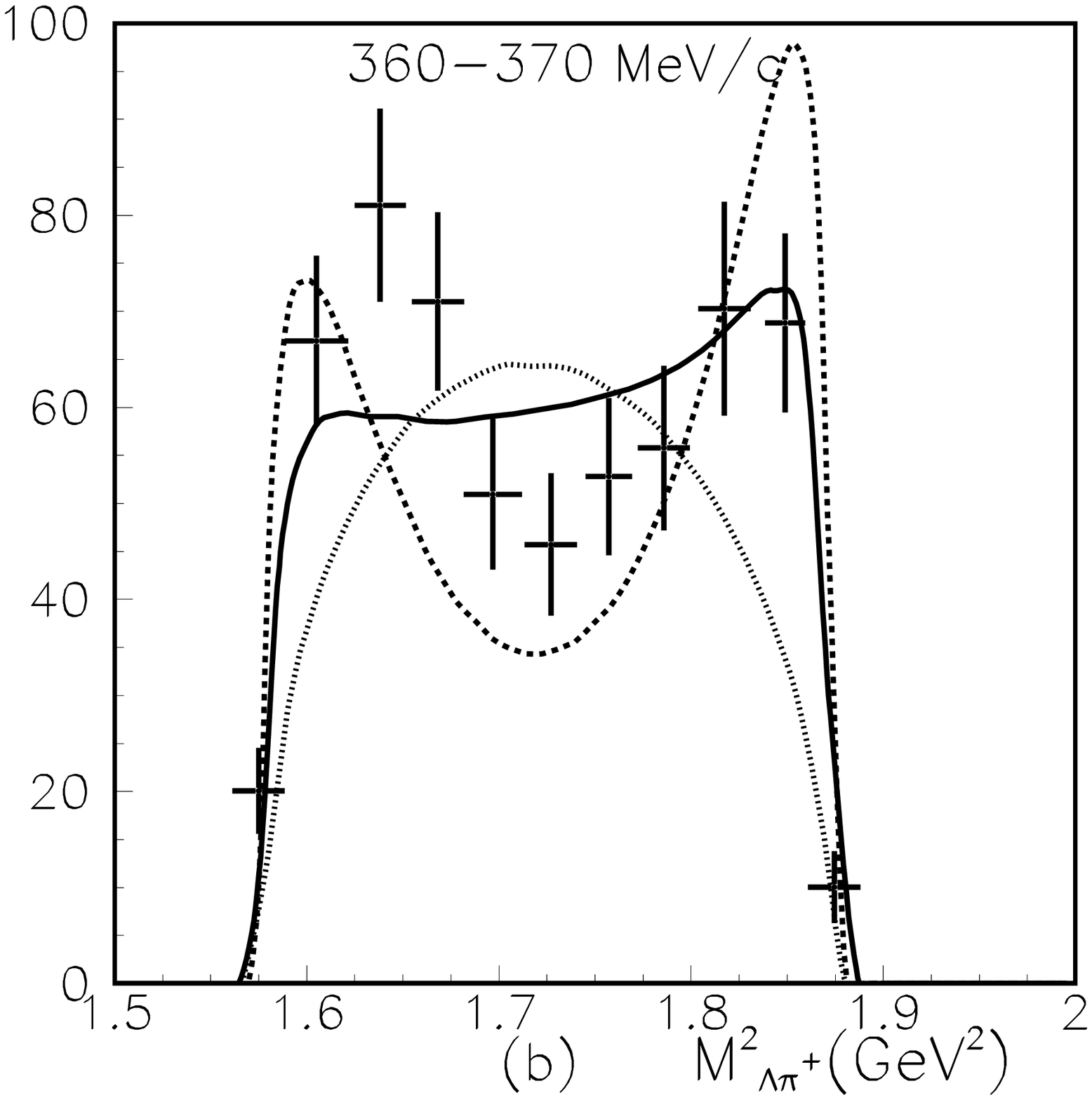}
\includegraphics[width=0.3\columnwidth]{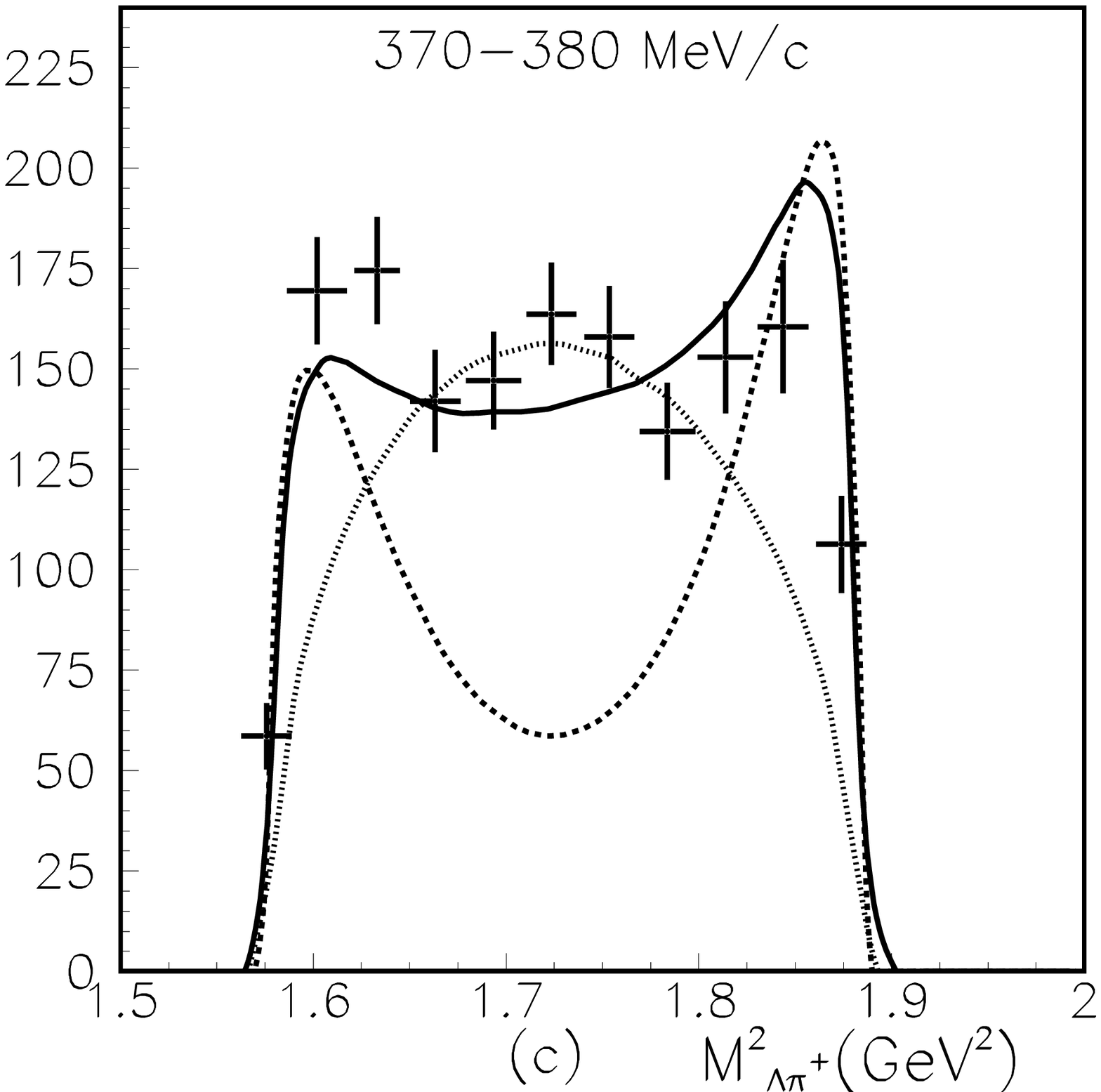}
\includegraphics[width=0.3\columnwidth]{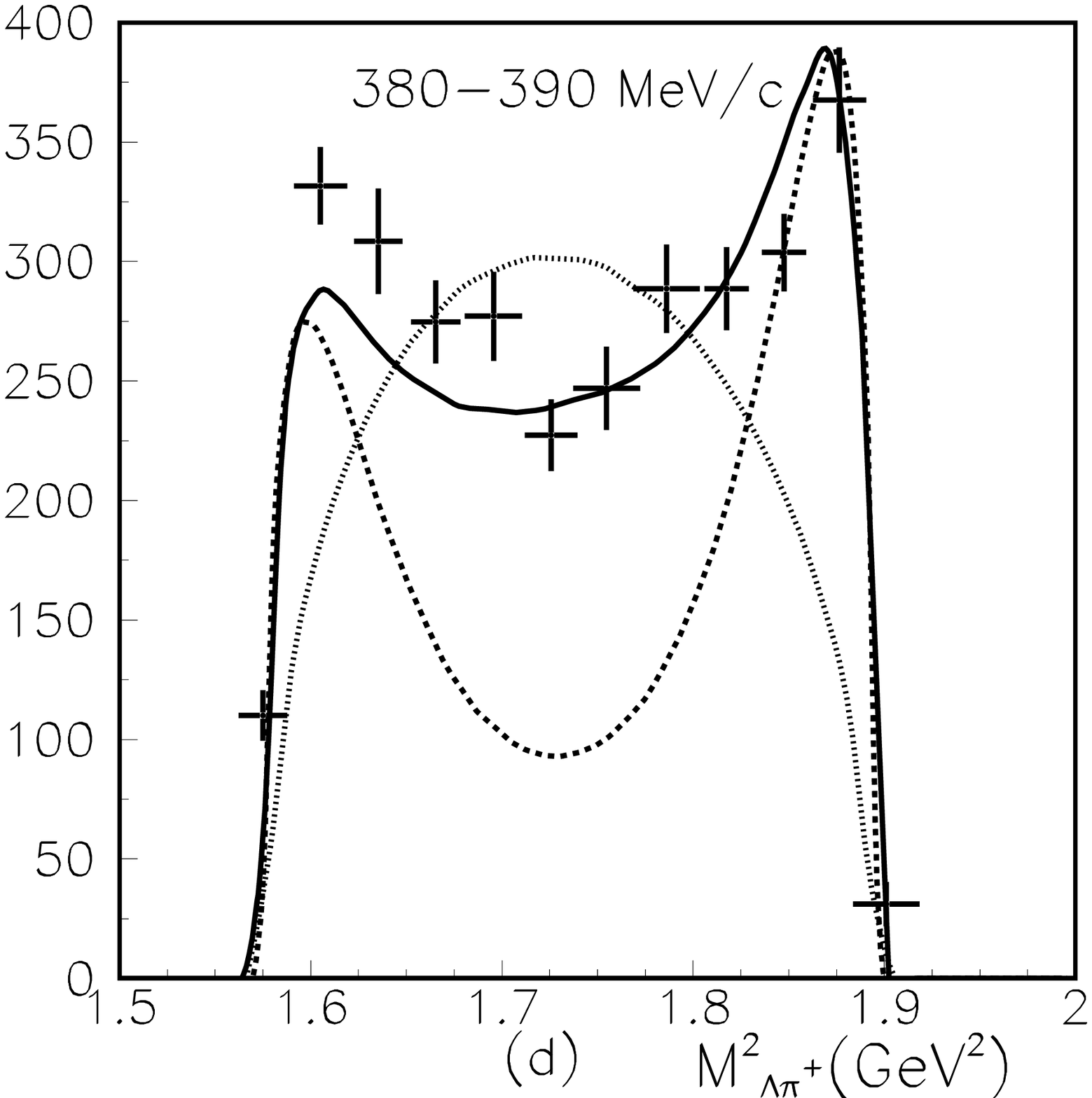}
\includegraphics[width=0.3\columnwidth]{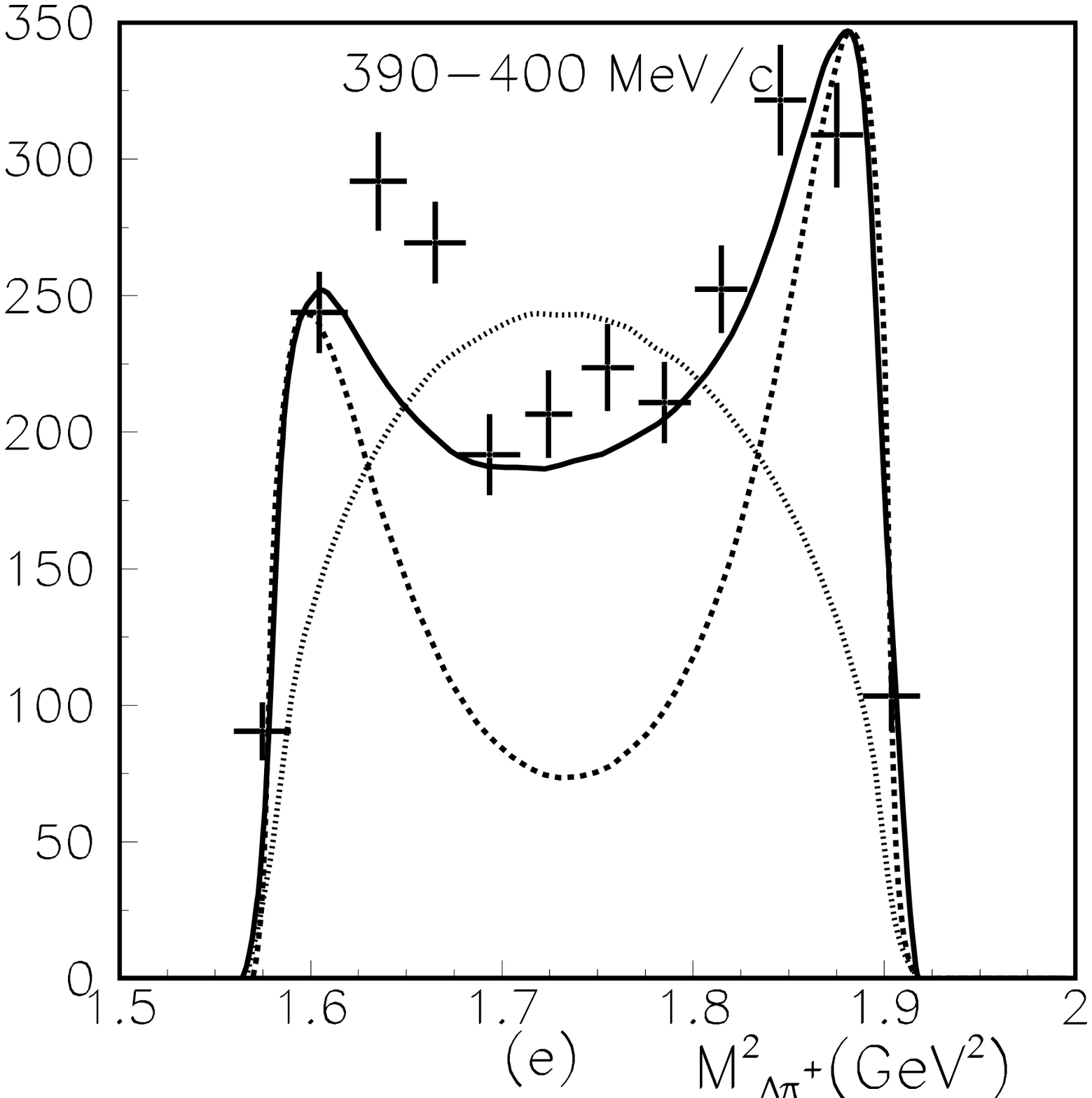}
\includegraphics[width=0.3\columnwidth]{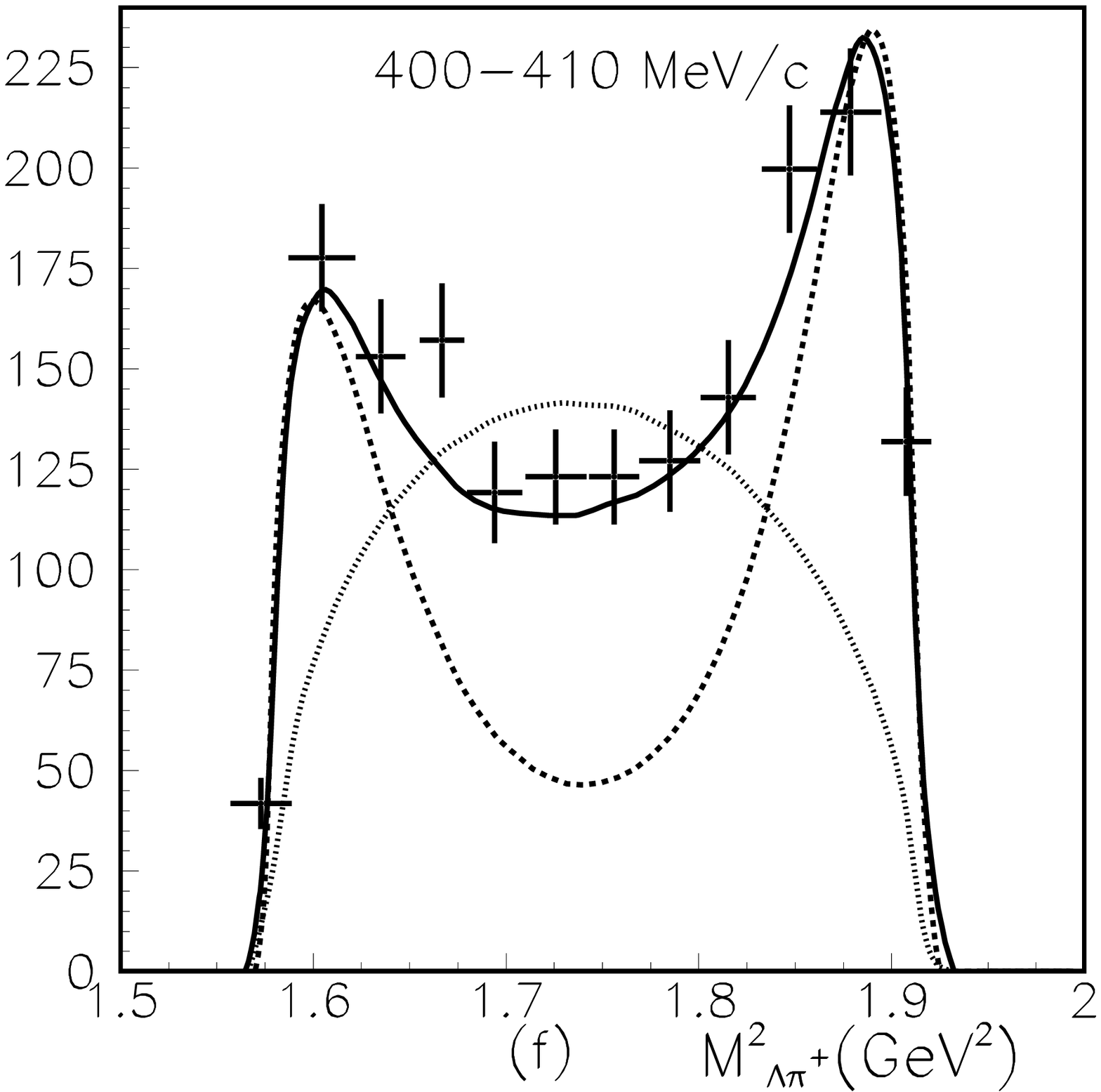}
\includegraphics[width=0.3\columnwidth]{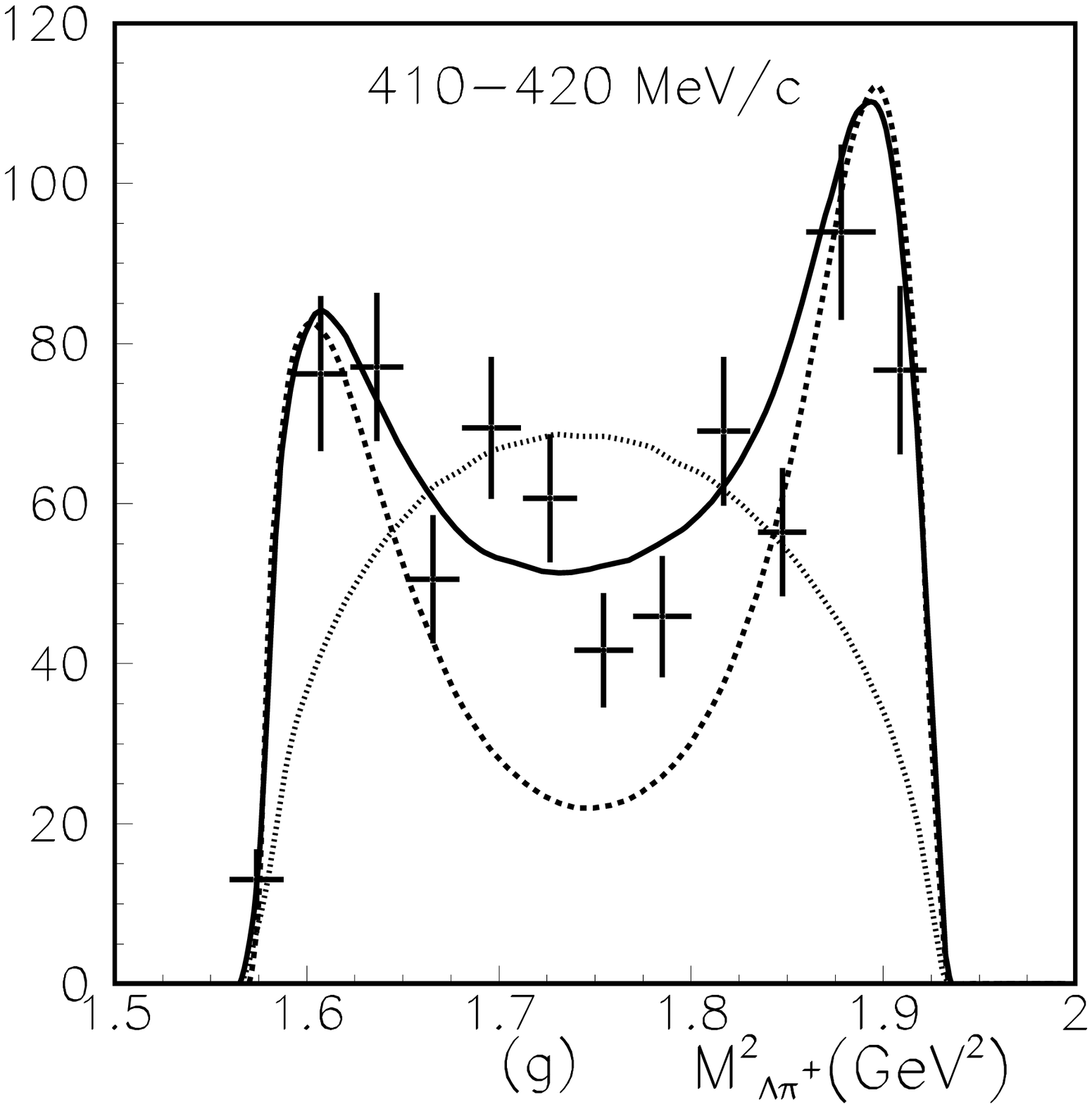}
\includegraphics[width=0.3\columnwidth]{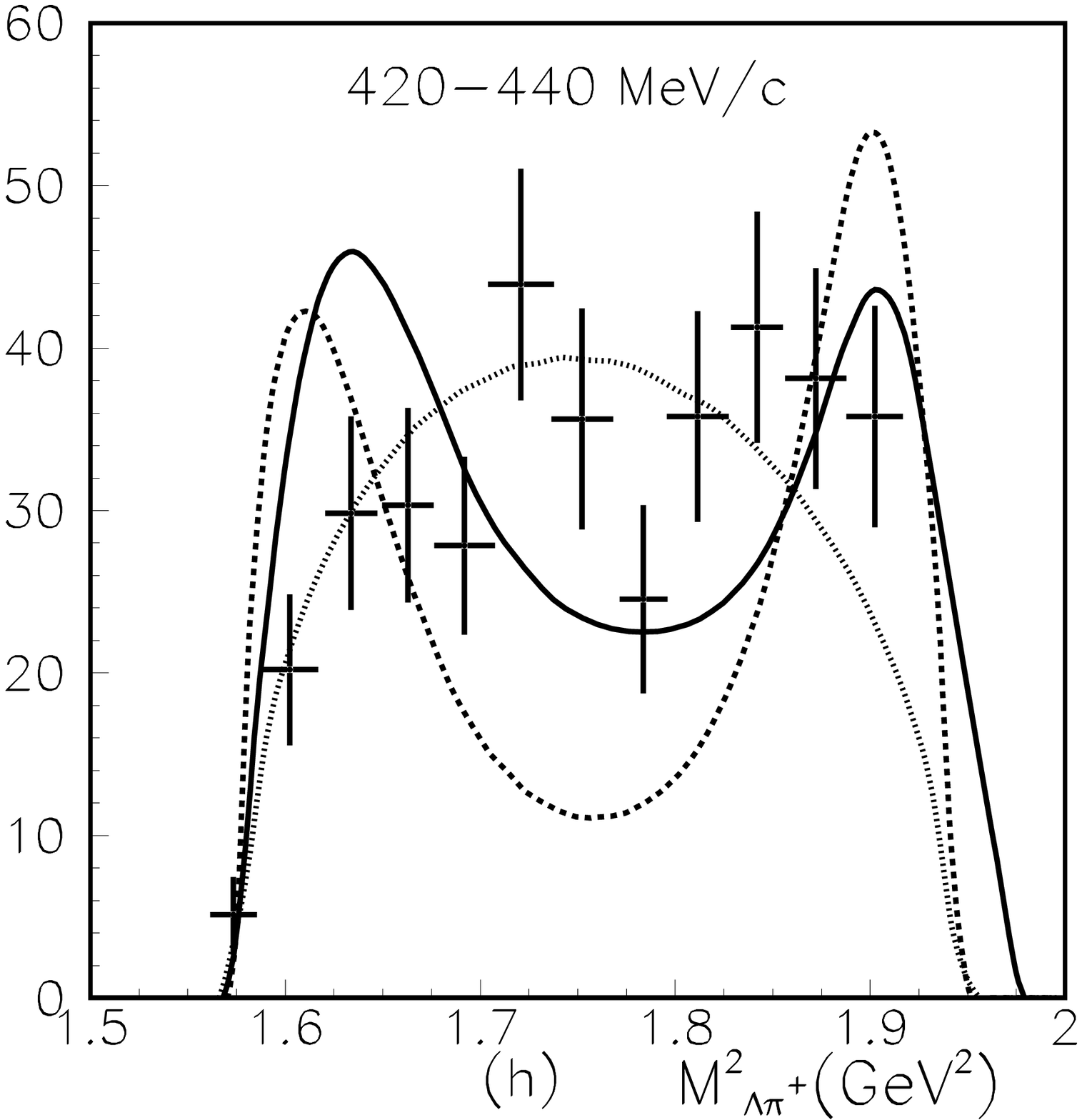}
\includegraphics[width=0.3\columnwidth]{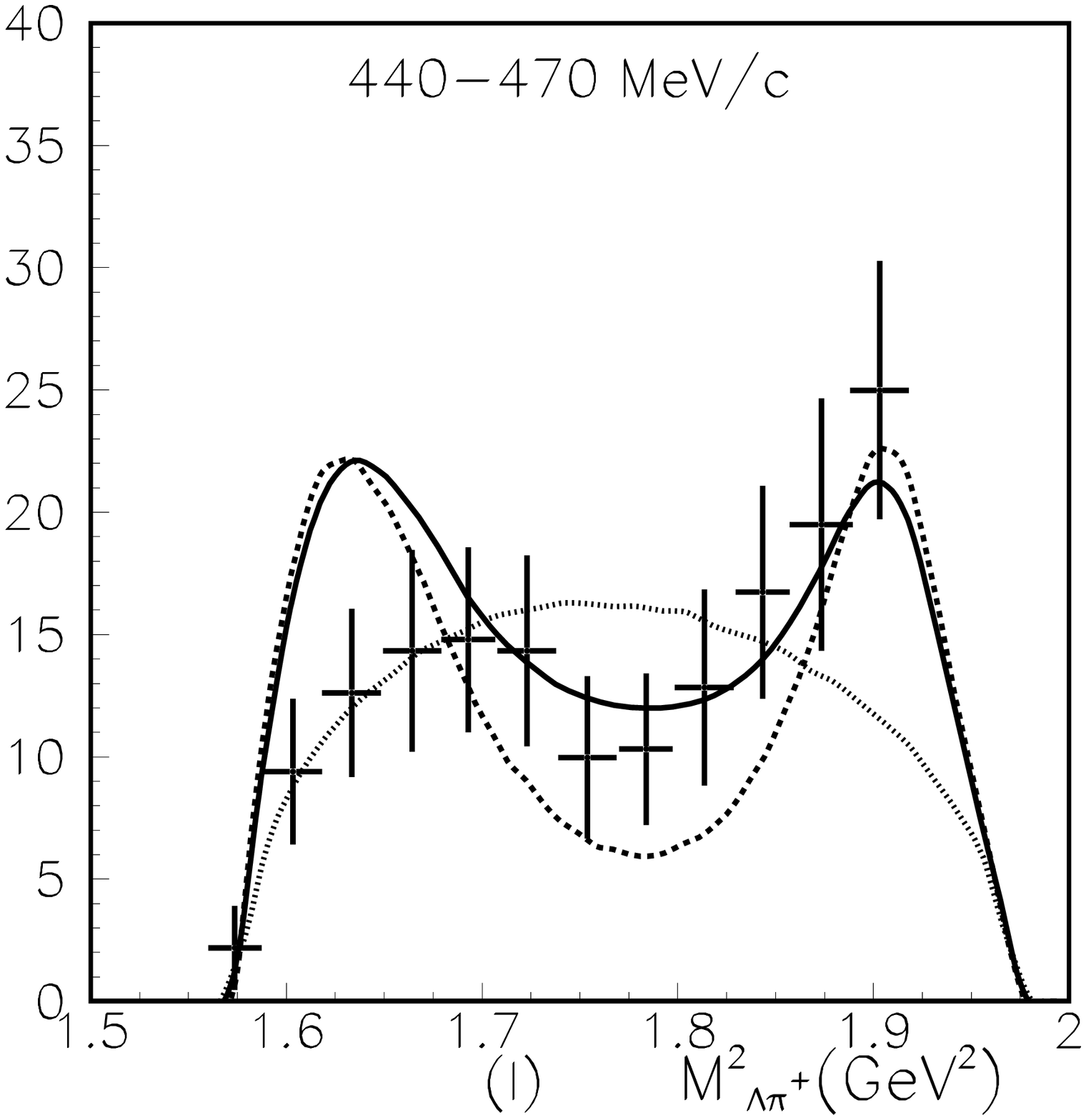}
\caption{Theoretical $\Lambda\pi^+$ invariant mass squared
distribution for various $K^-$ beam momenta compared with
data~\cite{exp}. The dotted line is for the pure $\Sigma^{*}(1385)$
with $J^P = {\frac{3}{2}}^+$; the solid line includes
$\Sigma^{*}_{1/2}$ in addition with $R_{3/2}=0.60$.} \label{lpip}
\end{center}
\end{figure}

\begin{figure}[htbp] \vspace{-0.cm}
\begin{center}
\includegraphics[width=0.3\columnwidth]{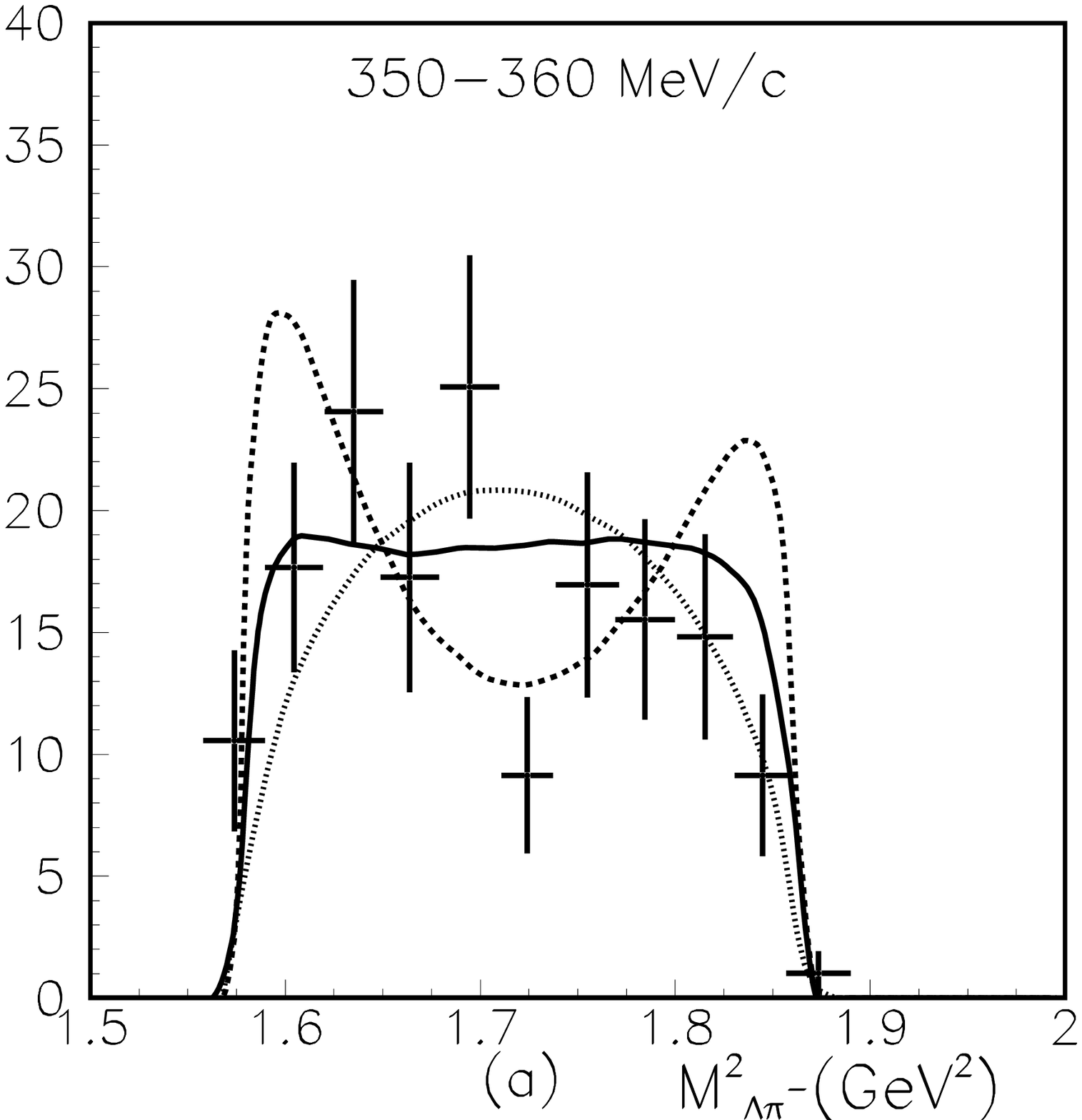}
\includegraphics[width=0.3\columnwidth]{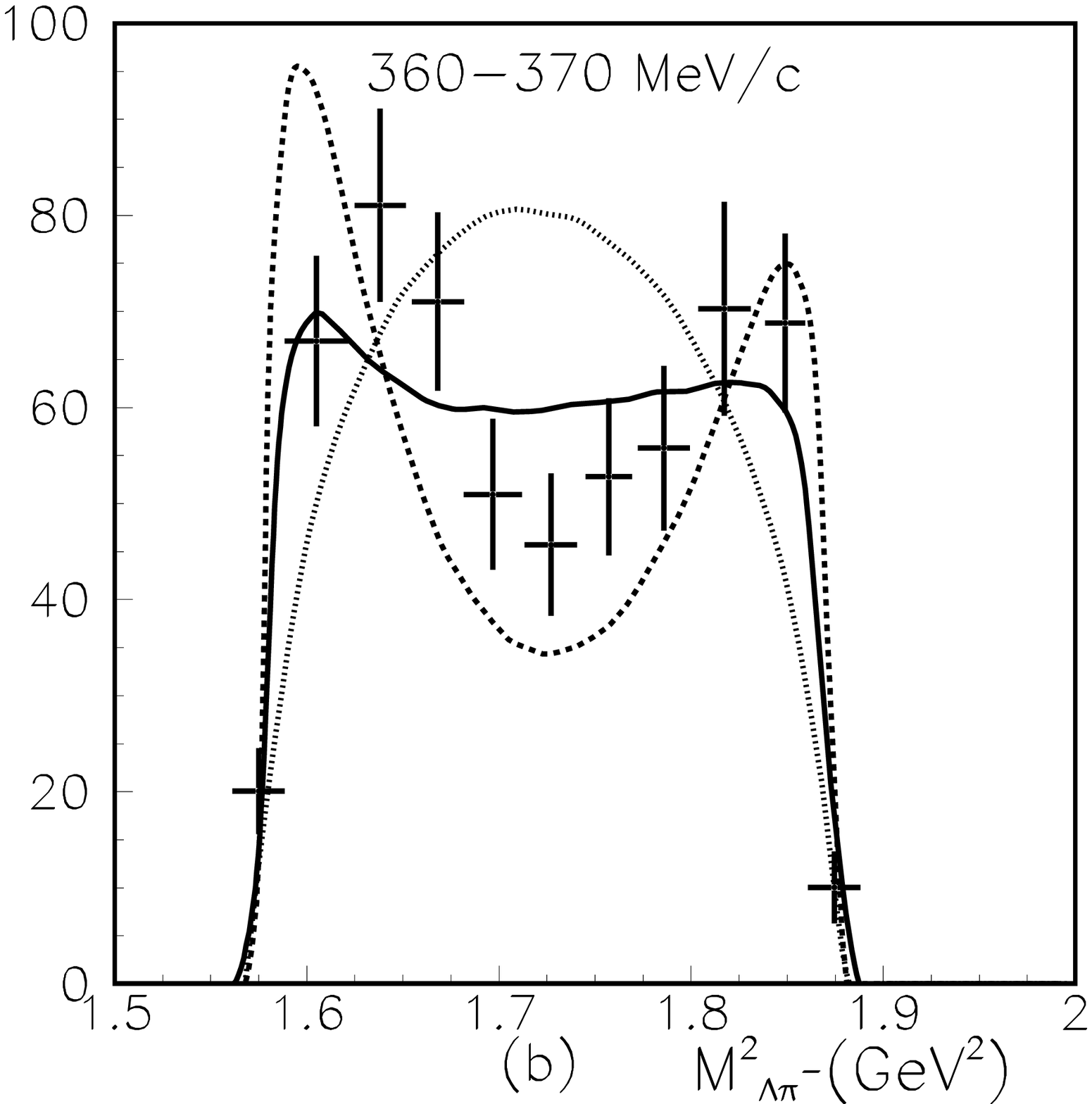}
\includegraphics[width=0.3\columnwidth]{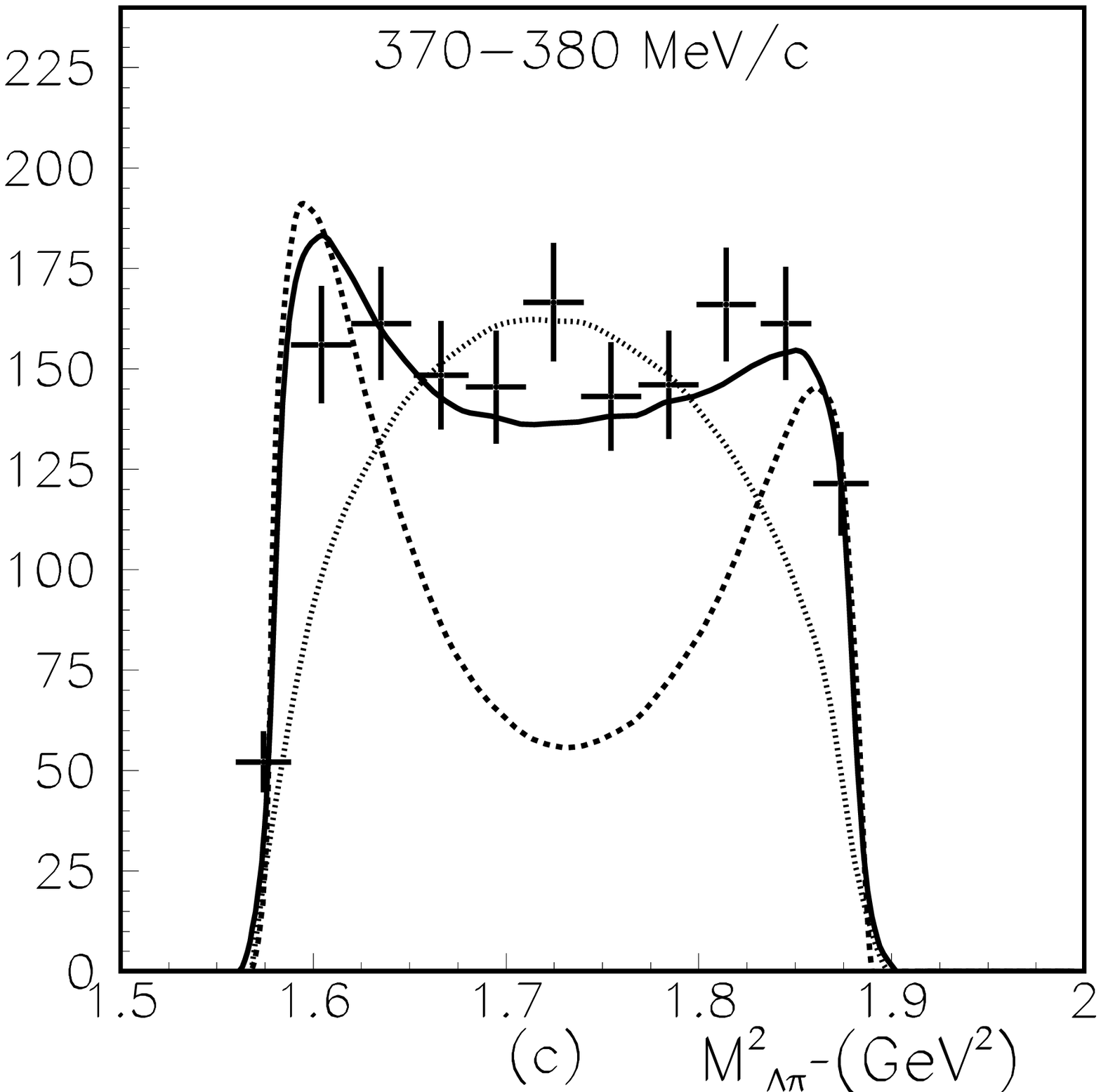}
\includegraphics[width=0.3\columnwidth]{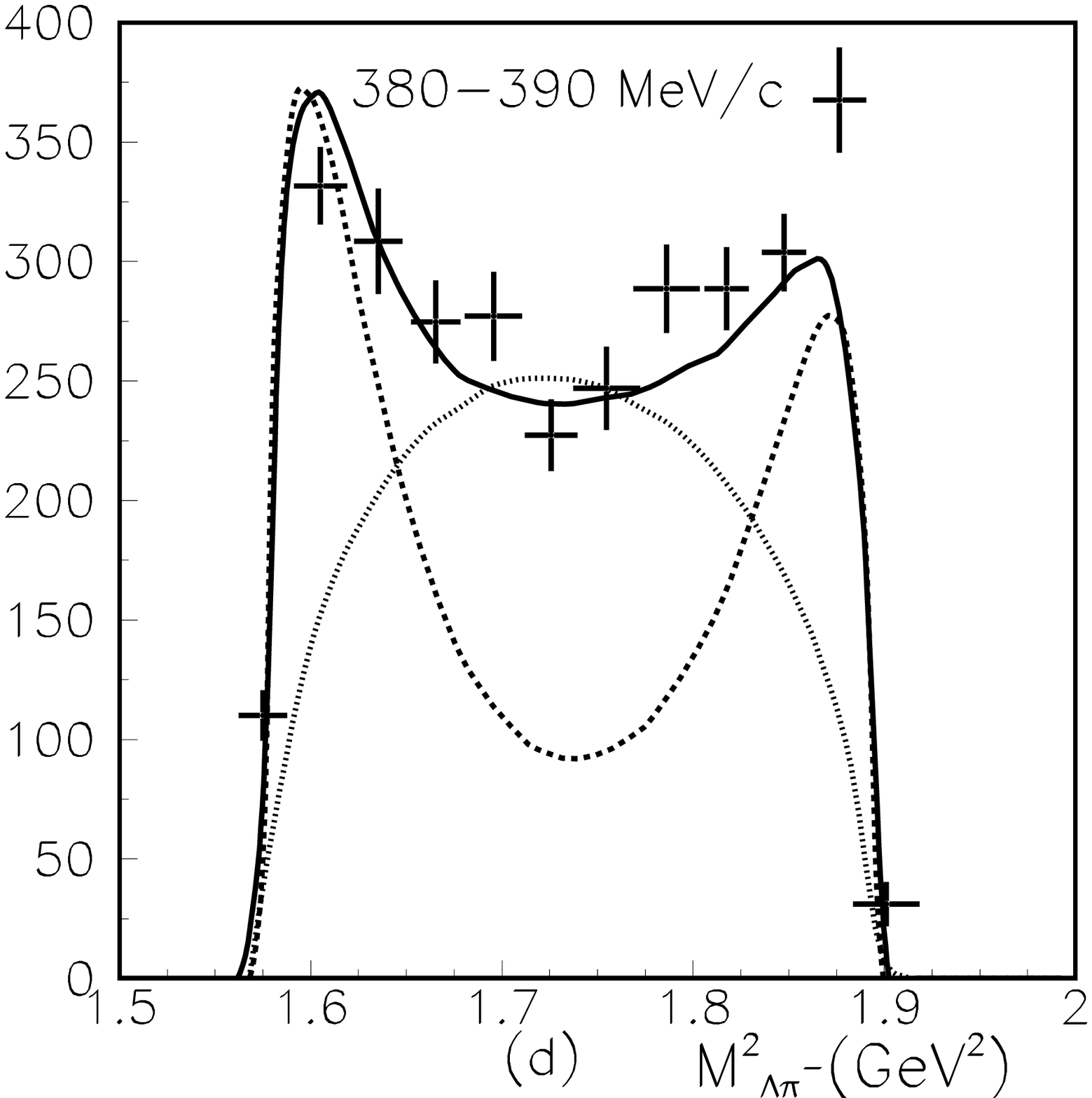}
\includegraphics[width=0.3\columnwidth]{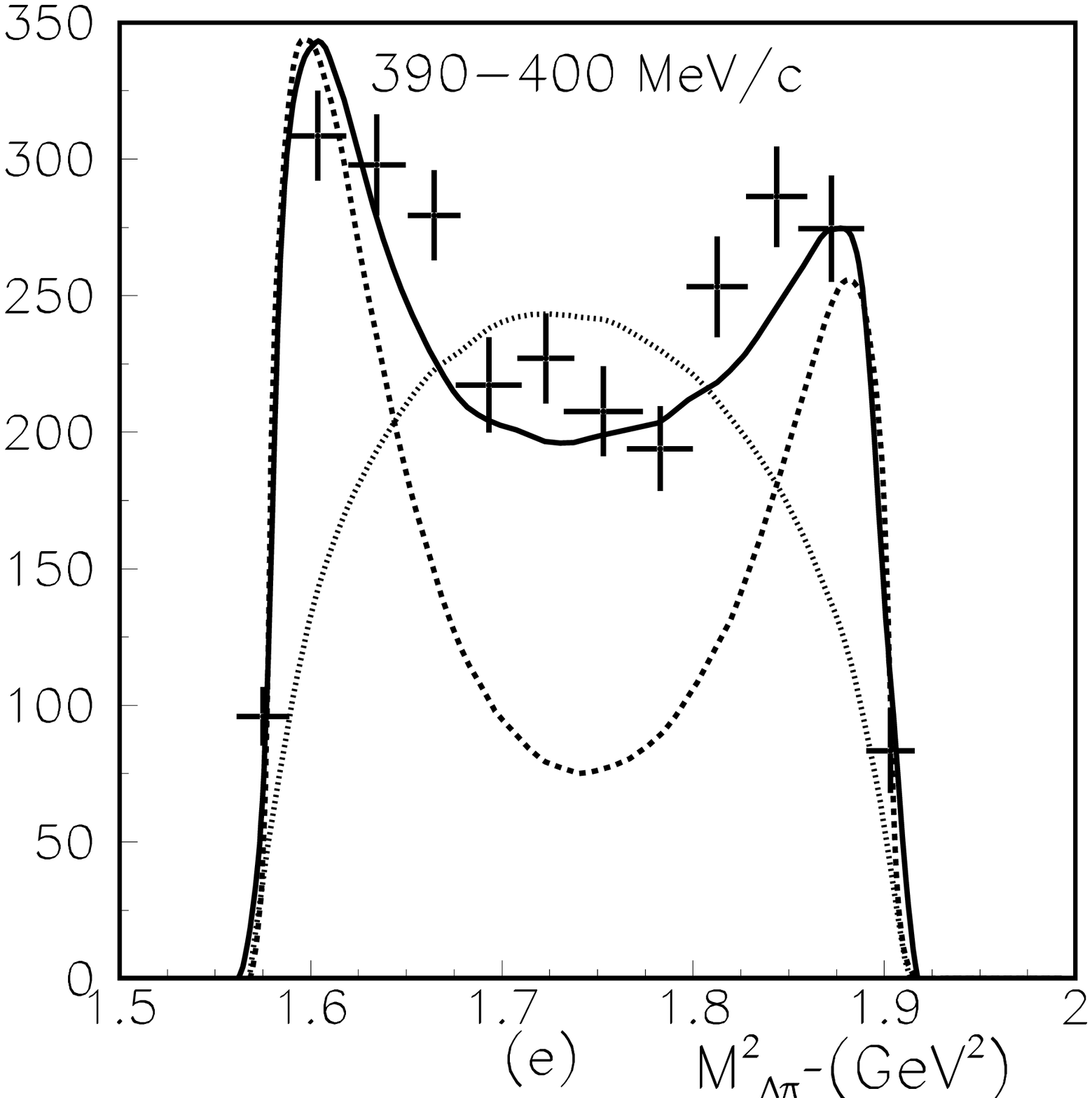}
\includegraphics[width=0.3\columnwidth]{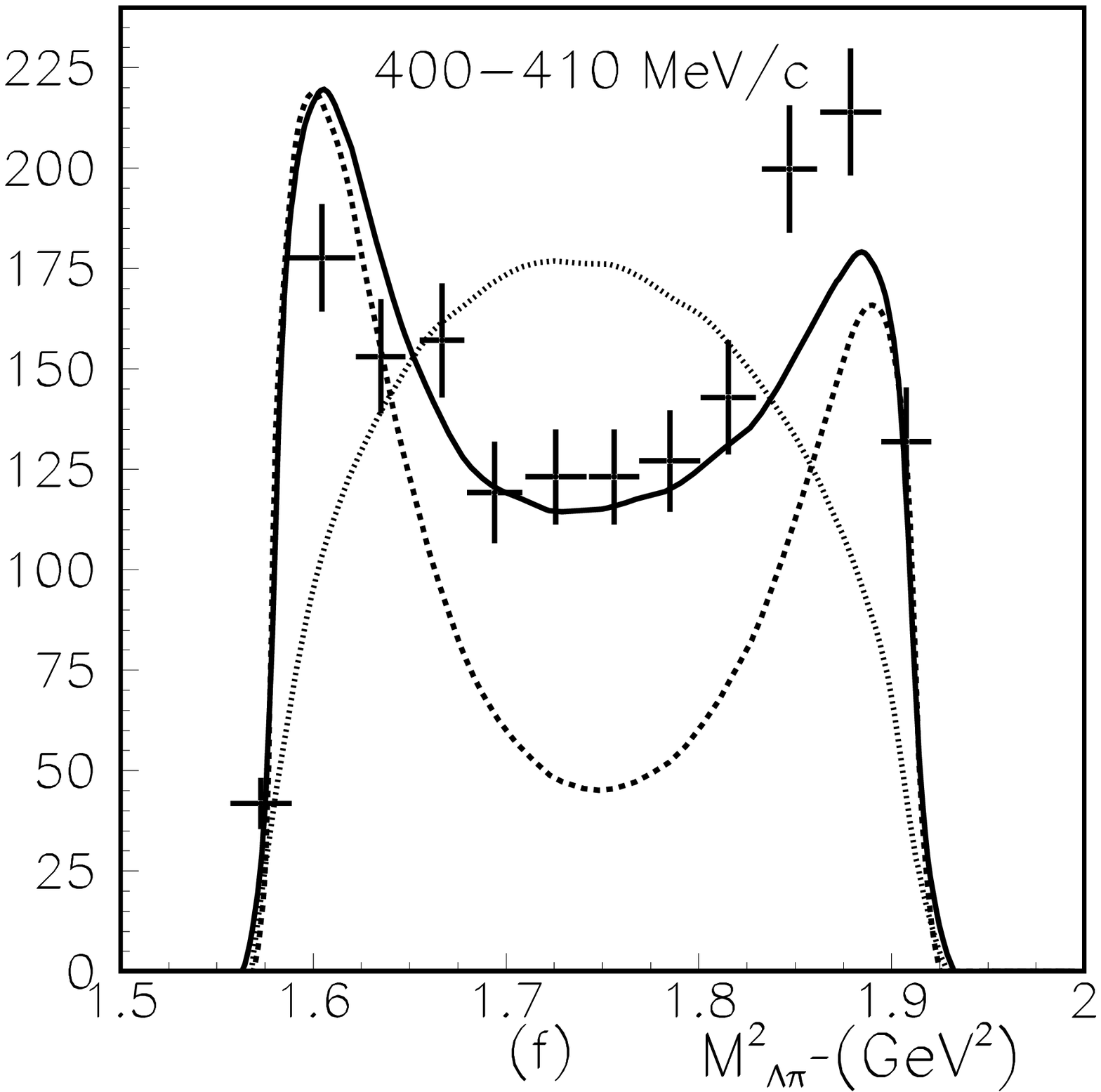}
\includegraphics[width=0.3\columnwidth]{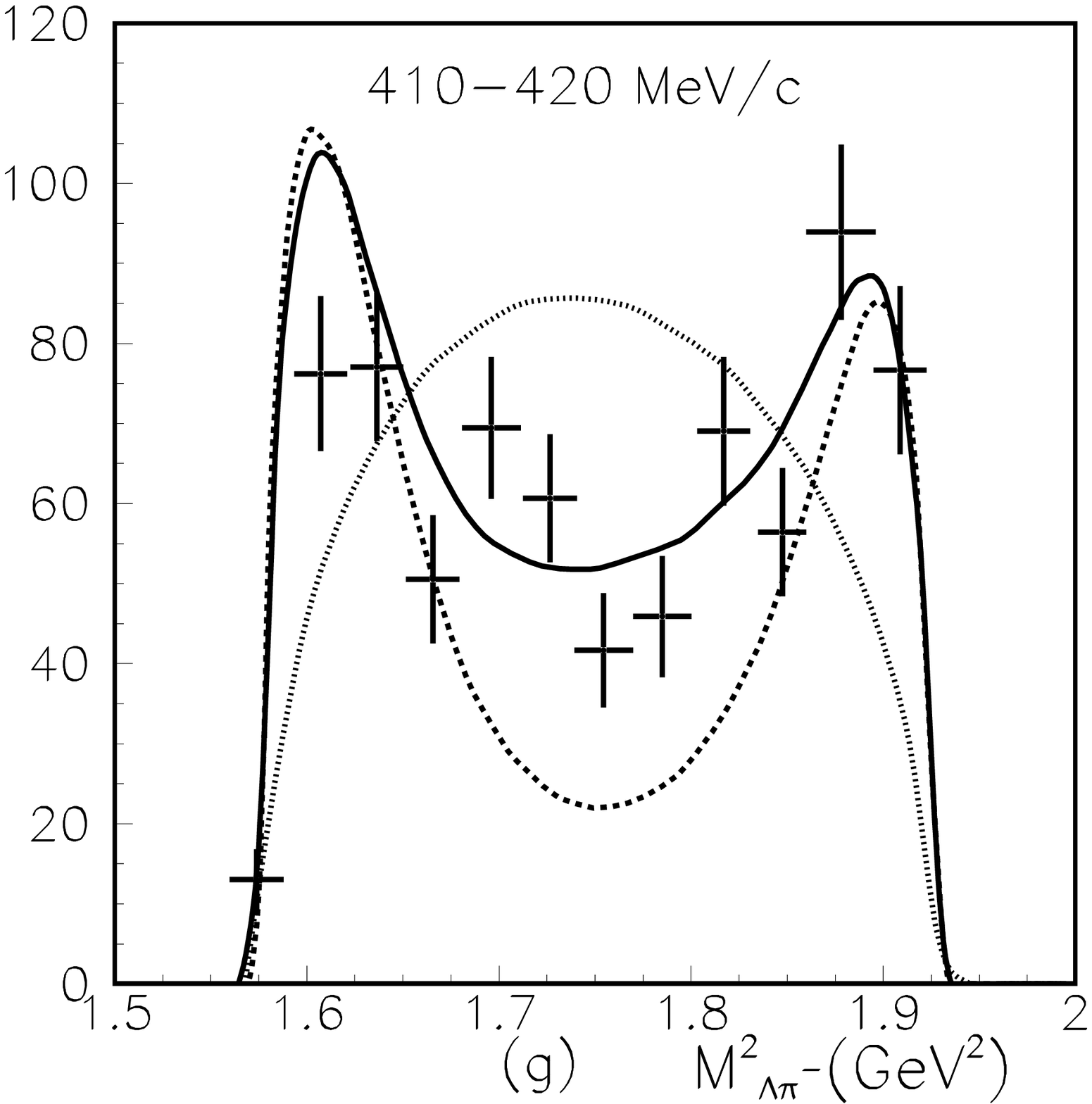}
\includegraphics[width=0.3\columnwidth]{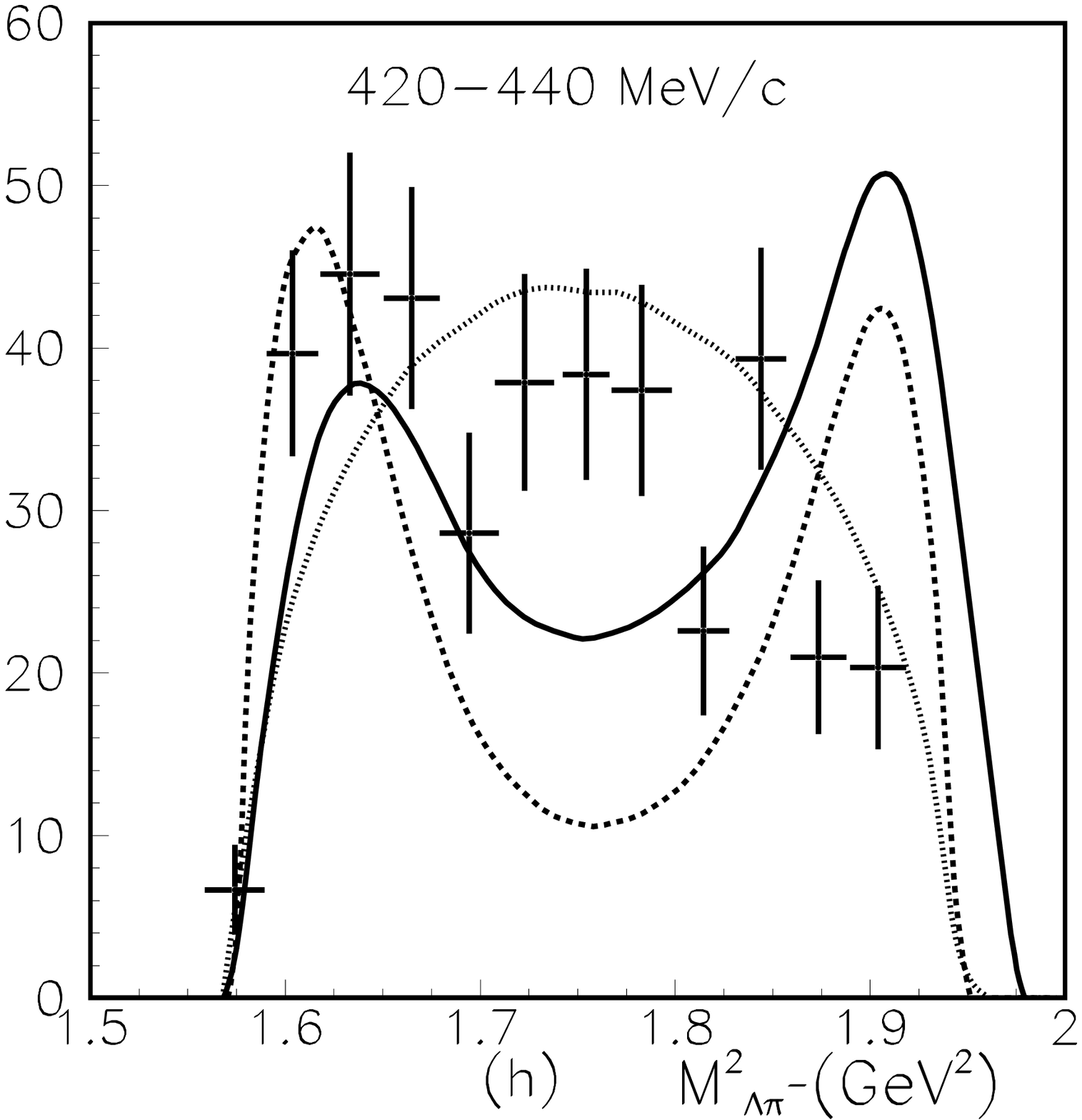}
\includegraphics[width=0.3\columnwidth]{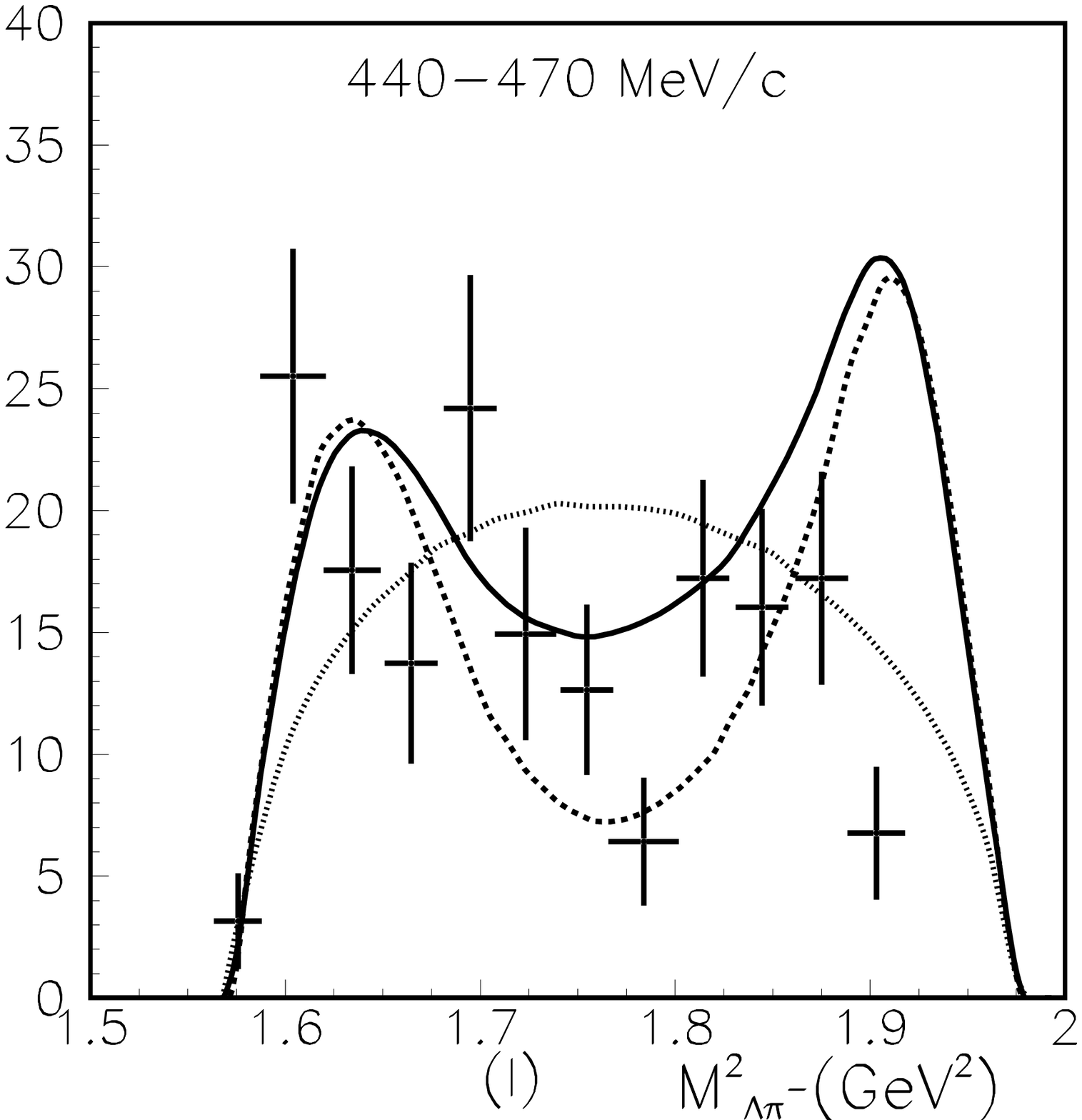}
\caption{Theoretical $\Lambda\pi^-$ invariant mass  squared
distribution for various $K^-$ beam momenta compared with
data~\cite{exp}. The dotted line is for the pure $\Sigma^{*}(1385)$
with $J^P = {\frac{3}{2}}^+$; the solid line includes
$\Sigma^{*}_{1/2}$ in addition with $R_{3/2}=0.60$.} \label{lpim}
\end{center}
\end{figure}

Comparison with the experimental data in Refs.\cite{exp,exp1,exp2}
for the total cross section of \kp\ in Fig.\ref{all} shows that for
the energies below 0.355GeV our theoretical calculation result does
not fit well with experiment. The reason may be the absence of the
contribution of $\Lambda^{*}(1405)$ or other resonance states. For
the energies larger than 0.42 GeV, the contribution of Feynman
diagram Fig.1(b) becomes large, but the contribution is uncertain
because of the large influence of form factor. Further detailed
study is necessary for this energy range. For the energies from
0.355 to 0.42 GeV, the theoretical prediction agrees very well with
the experiment, and the main contribution comes from the decay
$\Lambda^{*}(1520) \to \Sigma^{*\pm}\pi^{\mp}$. Therefore the decay
$\Lambda^{*}(1520) \to \Sigma^{*\pm}\pi^{\mp}$ is the interesting
place to search for the evidence of $\Sigma^{*}_{1/2}$.

The theoretical angular distributions in Fig.\ref{alam} of the
$\Lambda$ are almost the same for the pure $\Sigma^{*}(1385)$ with
$J^P = {\frac{3}{2}}^+$ and for the 60\% $\Sigma^{*}(1385)$ plus
40\% $\Sigma^*_{1/2-}$ ( $R_{3/2}=0.60$).

There are quite large differences between the two theoretical
invariant mass squared distributions which are shown in
Figs.\ref{pipi}, \ref{lpip}, \ref{lpim}, especially the invariant
mass squared spectra of $\Lambda\pi$. One may note that the solid
curves with both $\Sigma^{*}_{3/2}$ and $\Sigma^{*}_{1/2}$
contributions give much better agreement with the experiment data.

\begin{figure}[htbp] \vspace{-0.cm}
\begin{center}
\includegraphics[width=0.49\columnwidth]{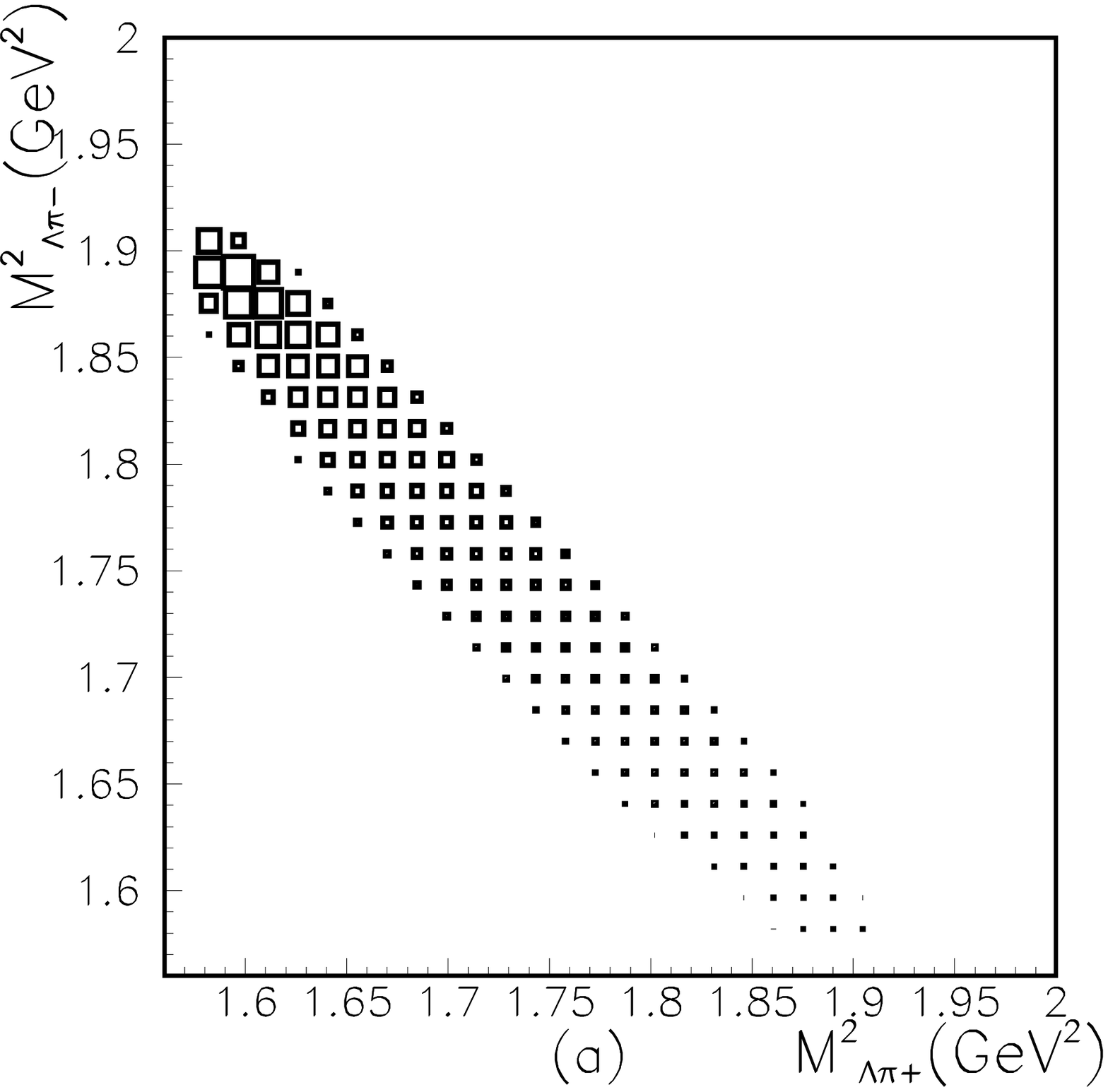}
\includegraphics[width=0.49\columnwidth]{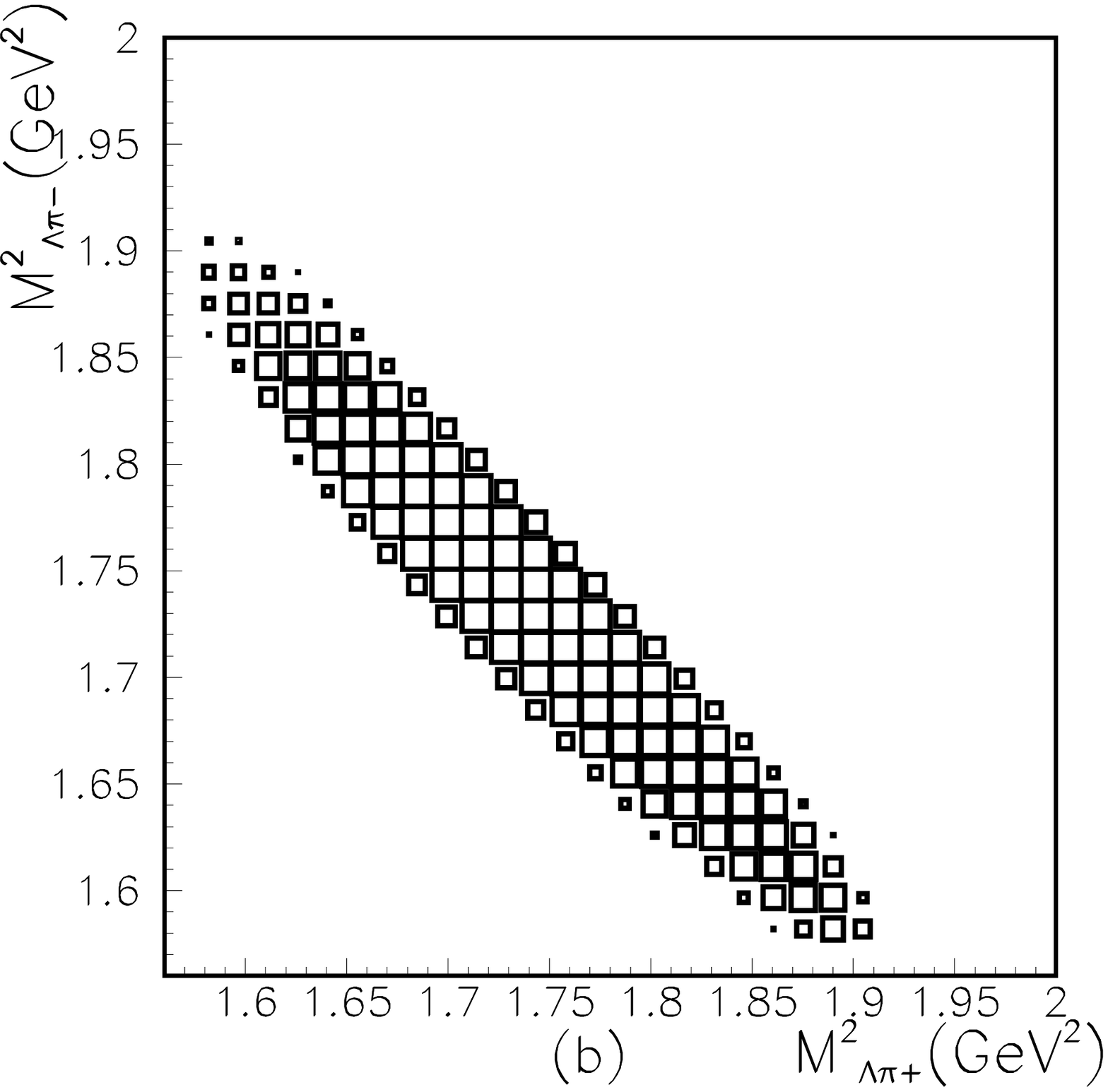}
\caption{ Dalitz plots (a) and (b) corresponding to the reactions
$K^{-}p \to \Lambda^{*} \to \Sigma^{*-}_{3/2}\pi^{+} \to
\Lambda\pi^{+}\pi^{-}$ and  $K^{-}p \to \Lambda^{*} \to
\Sigma^{*-}_{1/2}\pi^{+} \to \Lambda\pi^{+}\pi^{-}$, respectively,
at $Plab_{(K^{-})}=0.394$ GeV.} \label{da}
\end{center}
\end{figure}

To understand the reason for the difference, we show the Dalitz
plots for the two reaction sequences
\begin{eqnarray}
K^{-}p \to
\Lambda^{*} \to \Sigma^{*-}_{3/2}\pi^{+} \to
\Lambda\pi^{+}\pi^{-}\;\label{sig32}\\
K^{-}p \to \Lambda^{*} \to
\Sigma^{*-}_{1/2}\pi^{+} \to \Lambda\pi^{+}\pi^{-} \label{sig12}
\end{eqnarray}
at $Plab_{(K^{-})}=0.394GeV$ in Fig.\ref{da}. From Fig.\ref{da} (a)
and (b) we see that, the contribution of (\ref{sig32}) is
distributed on the top left corner, but of (\ref{sig12}) is in the
middle. This is because for the decay $\Lambda^{*} \rightarrow
\Sigma^{*-}_{3/2}\pi^{+}$, the final state particles are in the
relative S-wave, while for the decay $\Lambda^{*} \rightarrow
\Sigma^{*-}_{1/2}\pi^{+}$, they are in the relative P-wave with
large $\Sigma^{*-}_{1/2}$ width. From these Dalitz plots, one can
understand why there is so much difference in the invariant mass
squared spectra of $\Lambda\pi$. The experimental analysis in
Ref.\cite{exp} also considered the contribution of the S-wave of
$\Lambda\pi$, which may come from $\Sigma^{*}_{1/2}$,  also  the
$\pi^{+}\pi^{-}$ from the $\sigma$ or $\rho$, but the range of
invariance mass spectrum of $\pi^{+}\pi^{-}$ is from $0.28$ to $0.4$
GeV, which is far from the mass of $\sigma$ or $\rho$. By the
investigation we also find that the S-wave final state interaction
(FSI) of $\pi^{+}\pi^{-}$ has little influence on the $\Lambda\pi$
invariant mass squared spectra. Thus we conclude that there should
be contribution from the $\Sigma^{*}_{1/2}$ for the reaction at
energies around the $\Lambda^{*}(1520)$ peak. For the
$\pi^{+}\pi^{-}$ invariance mass spectra, the inclusion of $40\%$
$\Sigma^{*}_{1/2}$ gives some enhancement to the low energy end and
reproduces better the data for $Kp$ center of mass energies around
the $\Lambda^{*}(1520)$ peak.

In summary, we study the \kp\ reaction at $Plab_{(K^{-})}=0.25 -
0.60$ GeV. In our calculations we take into account all possible
form factors and final state interactions. We find that by including
$40\%$ $\Sigma^{*}_{1/2}$ contribution the theory agrees much better
with the experimental data~\cite{exp} for $Plab_{(K^{-})}$ in the
range of $0.355- 0.42$ GeV, corresponding to the $Kp$ center-of-mass
energies just under the $\Lambda^*(1520)$ peak. Through the
analysis, the difference between the two cases, with or without
$\Sigma^{*}_{1/2}$, comes from the different partial waves, namely,
S-wave of $\Sigma^{*}_{3/2}\pi$ verus P-wave of
$\Sigma^{*}_{1/2}\pi$ from $\Lambda^{*}(1520)$ decays, and the
different width of the two particles. The results of this work
strongly suggest that the new particle $\Sigma^{*}$ with
$J^P=\frac{1}{2}^{-}$ exists in the $\Lambda^{*}(1520) \to
\Sigma^{*}\pi\to\Lambda\pi\pi$ decays. Higher statistic data
experiments are necessary to establish this new resonance and to
understand its property.

\bigskip
\noindent {\bf Acknowledgements}  We thank Pu-ze Gao for useful
discussions. This work is partly supported by the National Natural
Science Foundation of China (NSFC) under grants Nos. 10875133,
10821063, 10635080, 10665001, and by the Chinese Academy of Sciences
under project No. KJCX3-SYW-N2, and by the Ministry of Science and
Technology of China (2009CB825200).

\end{document}